\documentclass{aa}
\usepackage[dvipsnames]{xcolor} 
\usepackage{graphicx,txfonts}
\usepackage[breaklinks=true]{hyperref} 
\hypersetup{colorlinks=true,citecolor=Blue}
\usepackage{natbib}
\bibpunct{(}{)}{;}{a}{}{,} 
\usepackage[english]{babel}
\usepackage{blindtext}
\usepackage{pifont}
\usepackage[normalem]{ulem}
\usepackage{comment}
\usepackage{placeins}
\usepackage{amsmath}
\newcommand{\PDS}{PDS\,70} 
\newcommand{\Mjup}{$M_{\rm{Jup}}$}

\begin{document}

   \title{Survival of the long-lived inner disk of \PDS}

   \author{Paola Pinilla
          \inst{1,2}
          \and
          Myriam Benisty\inst{3,4}
          \and
          Rens Waters \inst{5,6}
          \and
          Jaehan Bae \inst{7}
          \and
          Stefano Facchini \inst{8}
          }


   \institute{Mullard Space Science Laboratory, University College London, Holmbury St Mary, Dorking, Surrey RH5 6NT, UK\email{p.pinilla@ucl.ac.uk}
   \and
   Max-Planck-Institut f\"ur Astronomie, K\"onigstuhl 17, 69117, Heidelberg, Germany
    \and
    Universit\'{e} C\^{o}te d'Azur, Observatoire de la C\^{o}te d'Azur, CNRS, Laboratoire Lagrange, Bd de l'Observatoire, CS 34229, 06304 Nice cedex 4, France
    \and
    Universit\'{e} Grenoble Alpes, CNRS, Institut de Plan\'{e}tologie et d'Astrophysique (IPAG), F-38000 Grenoble, France
    \and
    Department of Astrophysics/IMAPP, Radboud University, PO Box 9010, 6500 GL Nijmegen, The Netherlands
    \and
    SRON, Niels Bohrweg 2, Leiden, The Netherlands
    \and
    Department of Astronomy, University of Florida, Gainesville, FL 32611, USA
    \and
    Dipartimento di Fisica, Universit\`a degli Studi di Milano, Via Celoria 16, 20133 Milano, Italy
              }
 
  \abstract
  {The K7 T Tauri star\PDS~remains the best laboratory for investigating the influence of giant planet formation on the structure of the parental disk. One of the most intriguing discoveries is the detection of a resolved inner disk from ALMA observations that extends up to the orbit of \PDS b. It is challenging to explain this inner disk because most of the dust particles are expected to be trapped at the outer edge of the gap opened by \PDS b and \PDS c. By performing dust evolution models in combination with radiative transfer simulations that match the gas disk masses obtained from recent thermo-chemical models of \PDS, we find that when the minimum grain size in the models is larger than 0.1$\mu$m, there is an efficient filtration of dust particles, and the inner disk is depleted during the first million year of dust evolution. To maintain an inner disk, the minimum grain size in the models therefore needs to be smaller than 0.1$\mu$m. Only when grains are that small are they diffused and dragged along with the gas throughout the gap opened by the planets. The small grains transported in the inner disk grow and drift into it, but the constant reservoir of dust particles that are trapped at the outer edge of the gap and that continuously fragment allows the inner disk to refill on million-year timescales. Our flux predictions at millimeter wavelength of these models agree with ALMA observations. These models predict a spectral index of 3.2 in the outer and 3.6 in the inner disk. Our simple analytical calculations show that the water emission in the inner disk that was recently observed with the James Webb Space Telescope may originate from these ice-coated small grains that flow through the gap, grow, and drift toward the innermost disk regions to reach the water snowline. These models may mirror the history and evolution of our Solar System, in which Jupiter and Saturn played a crucial role in shaping the architecture and properties of the planets.}
  
   \keywords{accretion, accretion disks – circumstellar matter – planets and satellites: formation – protoplanetary disks}

   \authorrunning{P.~Pinilla et al.}
   \maketitle
%

\section{Introduction} \label{sect:intro}

Although planets form in protoplanetary disks, it remains hard to discover them while they are still surrounded by their parental disk remains. Protoplanet candidates are still highly debated \citep[][]{asensio2021, benisty2022, Ren2023}. The K7 T Tauri star \PDS~remains the best laboratory for studying the influence of giant planet formation on the disk structure. 

\PDS~is 5.4\,Myr old and is located in the  Upper Centaurus Lupus association at a distance of 113.47\,pc \citep{muller2018, gaia2021edr3}. It hosts a disk with a large cavity that is visible at different wavelengths, {including scattered light  \citep{dong2012} and (sub-) millimeter  observations \citep{hashimoto2015, keppler2019}}. Inside the cavity, two planets at separations of $\sim22$\,au (\PDS b) and $\sim34$\,au (\PDS c) have been discovered \citep{keppler2018, muller2018} that are still accreting material from the disk \citep[with values of $\sim10^{-8}-10^{-7}M_{\rm{Jup}}\,\rm{yr}^{-1}$ for \PDS b and $\sim10^{-8}M_{\rm{Jup}}\,\rm{yr}^{-1}$ for \PDS c,][]{aoyama2019,haffert2019, Thanathibodee2020, hashimoto2020, zhou2021}. The masses of these planets are difficult to constrain observationally. {\cite{wang2021} provided dynamical constrains of the planet masses by enforcing dynamical stability of the planet orbits. They obtained values of $3.2^{+3.3}_{-1.6}\,M_{\rm{Jup}}$ for \PDS b and $7.5^{+4.7}_{-4.2}\,M_{\rm{Jup}}$ for \PDS c.} Most likely, the masses of \PDS b and \PDS c range between 1-10\,\Mjup, which is also supported  by hydrodynamical models and radiative transfer simulations of this system \citep{bae2019, toci2020}.

The planetary accretion of \PDS b and \PDS c suggests that gas flows through the gap. This has been supported by recent observations from The Atacama Large Millimeter/submillimeter Array (ALMA) of different molecules  \citep{keppler2019, facchini2021}, where $^{12}$CO and HCO$^{+}$ were observed  within the orbit of \PDS b. {\cite{portilla2023} analyzed these ALMA observations and suggested that the depletion factor of the gas density inside the gap is such that both planets are as massive as $4\,M_{\rm{Jup}}$}. Circumplanetary disks (CPDs) around both planets have been directly or indirectly observed. For \PDS b, the spectrum in the K band from the Very Large Telescope/SINFONI supports a scenario of dust around \PDS b, which was interpreted as a CPD \citep{Christiaens2019}. The CPD of  \PDS c was directly detected with ALMA \citep{isella2019, benisty2021}. Interestingly, tentative co-orbital submillimeter emission within the Lagrangian region L5 of the protoplanet \PDS b has recently been  suggested by reanalyzing the ALMA data \citep{balsalobre2023}.

One of the most intriguing discoveries from the ALMA observations is the detection of a resolved inner disk \citep{long2018_pds, keppler2019, benisty2021}. This is consistent with the detection of near-infrared excess in the spectral energy distribution \citep[SED,][]{dong2012}, which could be explained by a population of small (micron-sized) dust particles close to the star.  It is currently unclear how this inner disk can exist and be sustained over million-year timescales with the current estimate of mass accretion rate onto the star \citep{Campbell2023}. Due to the dust trapping at the outer edge of the common gap created by the two giant planets (\PDS b and \PDS c), it is expected that most of the dust is blocked and stops the dust filtration and flow of dust from the outer to the inner disk \citep[e.g.,][]{pinilla2016, joanna2019}. It is important to determine the physical conditions for which this inner disk can be sustained for understanding whether terrestrial planets can still form within this inner region. In addition, the pebble flux from the outer to the inner disk can also determine the number of essential elements for sustaining life, such as water, that could be available within the snowline in the gas phase and that may be accreted by forming planets. 

Observational evidence recently showed that water vapor is likely delivered to the inner disk through icy pebbles that drift inward \citep[e.g.,][]{salyk2011, salyk2019}. Observations of the luminosity of infrared H$_2$O emission from Spitzer and the James Webb Space Telescope (JWST) spectra and the spatially resolved dust disk radius obtained from ALMA images suggest that the disks that are large and host substructures (which retain the pebbles in the outer disk) with a lower H$_2$O content \citep{banzatti2020, banzatti2023}. This is supported by models that include dust evolution, substructures, water sublimation, and the diffusion of water vapor in the inner disk \citep[e.g.,][]{kalyaan2021, kalyaan2023}. This introduces a new idea in the field, according to which the origin of water on the inner terrestrial planets does not only rely on local collisions of water-bearing planetesimals, but may mainly come from drifting icy-particles.

The JWST observations of \PDS~with the Mid-InfraRed Instrument (MIRI)  revealed a wealth of water lines, implying the presence of water in the inner disk ($<1\,$au) and indicating that  potential  planets forming in the inner disk may have access to a water reservoir despite the two giant planets in the outer disk \citep{perotti2023}. This challenges current models of dust evolution with giant embedded planets that are massive enough to open deep gaps, such as \PDS b and \PDS c, where dust trapping is expected to be very efficient at the edge of their gap. 

In this paper, we perform dust evolution and radiative transfer models for the conditions of \PDS~ (Sect.~\ref{sect:models}) to investigate how its inner disk can survive over million-year timescales. The results of these models and the comparison with current and future ALMA observations are presented in Sect.~\ref{sect:results}. The results and limitations are discussed in Sect.~\ref{sect:discussion}, in addition to the comparison with  recent JWST observations. Finally, we present the main conclusions of this work in Sect.~\ref{sect:conclusions}.

\section{Models} \label{sect:models}

\subsection{Hydrodynamical models}

The results from hydrodynamical models of planet-disk interaction performed by \cite{bae2019} were used as input for the dust evolution models presented in this work. The models from \cite{bae2019} were tailored to simulate the structure of \PDS. The authors performed these simulations with the two-dimensional locally isothermal hydrodynamical code \texttt{FARGO-ADSG} \citep{Baruteau2019}, which is an extension of the publicly available \texttt{FARGO} code \citep{Benitez2016}. In the models taken for this work, the mass of \PDS b was 5\Mjup~and that of \PDS c was 2.5\Mjup. The disk viscosity ($\alpha$) was assumed to be constant  over radius as well as over time, and to be equal to $1\times10^{-3}$. The initial disk mass was $3\times10^{-3}$\,$M_\odot$.

To input the gas density distribution from the 2D (radial and azimuthal) hydrodynamical simulations to the 1D (radial) dust evolution models, the results from the hydro models were azimuthally averaged over 0.1\,Myr (between $t =$ 0.9 and 1 Myr) after the disk had reached a quasi-steady state \citep[Fig~2 in][]{bae2019}. After this time, the disk mass was $9\times10^{-4}$\,$M_\odot$, which is  the disk mass we used as input in  the dust evolution models.
The planets migrated in the simulations, which means that their radial locations changed over time, although the migration was pretty slow (0.3 and 0.5\,au\,Myr$^{-1}$ for b and c, respectively). 

The averaged gas surface density and radial grid were used for the dust evolution models, and we assumed that the gas density remained steady during the dust evolution described in the next subsection.

\begin{figure*}
\centering
    \tabcolsep=0.05cm 
    \begin{tabular}{ccc}  
     \includegraphics[width=6cm]{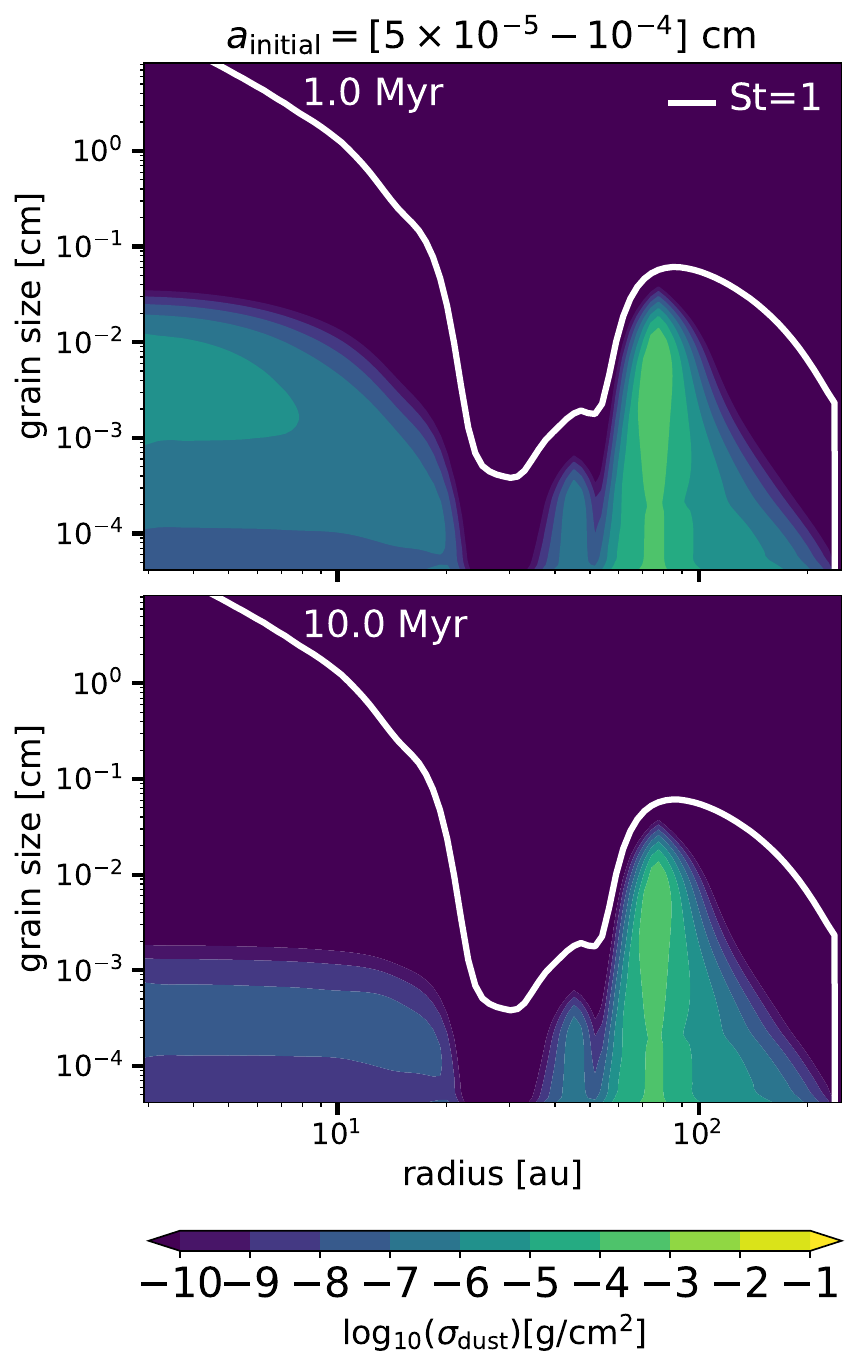}&
     \includegraphics[width=6cm]{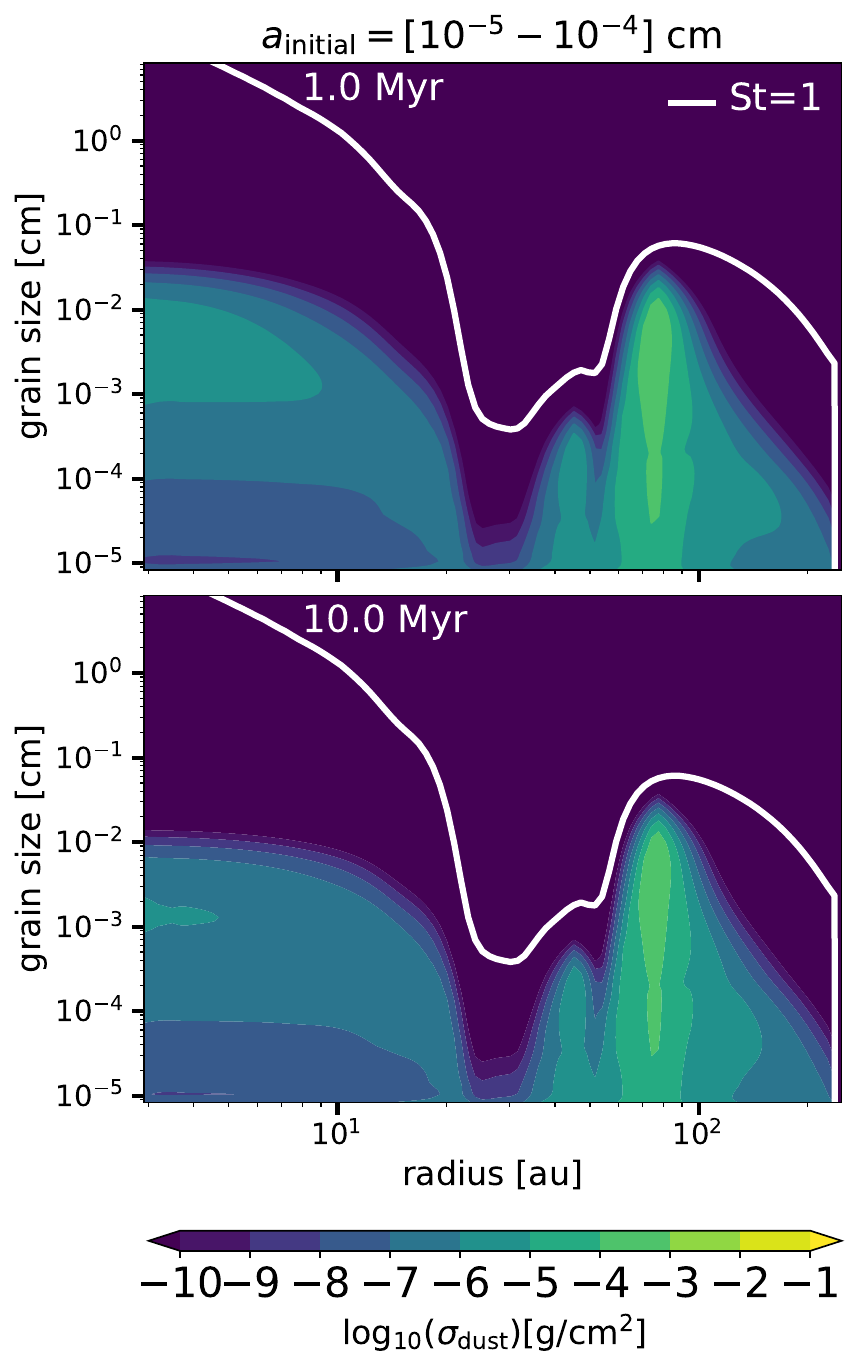}&
     \includegraphics[width=6cm]{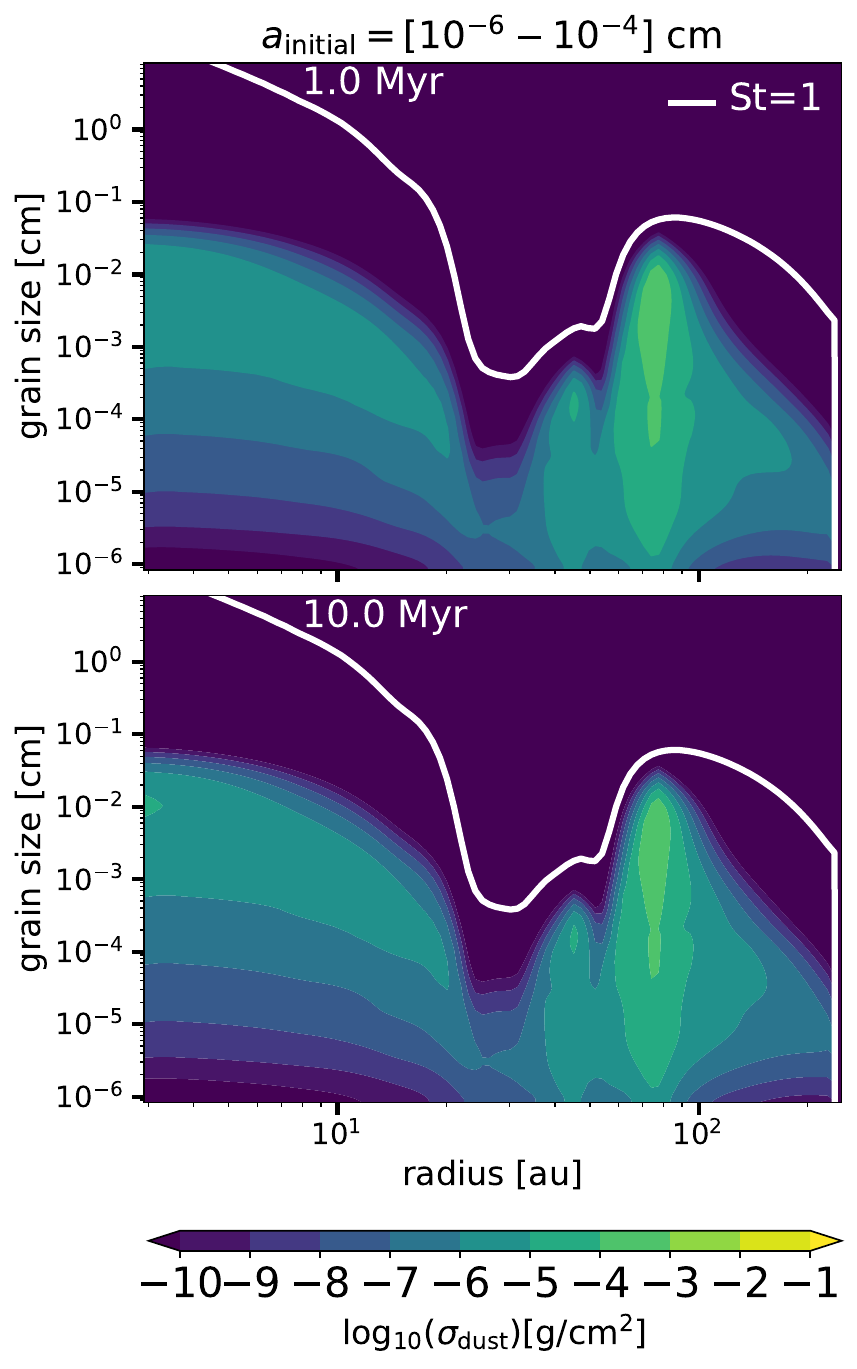}
    \end{tabular}
\caption{Results from dust evolution models for the \PDS~disk. Specifically, dust density distributions after 1\,Myr (top panels) and 10\,Myr (bottom panels) of evolution as a function of grain size (y-axis) and distance from the star (x-axis). The difference in simulations is the minimum grain size in the initial dust size distribution, that is, decreasing from the left to the right column. The y-axis  is different for each column because  the grain size grid depends on the initial grain size that is assumed (given in the title of each column). }
\label{fig:dust_evolution_models}
\end{figure*}

\subsection{Dust evolution models}

For the dust evolution, we used the code \texttt{Dustpy} \citep{stammler2022}, version 1.0.3. \texttt{Dustpy} calculates the growth and fragmentation of dust particles, as well as their dynamics. Because we assumed that the gas surface density profile remained constant over time, we did not include gas evolution in these models. Details about the equations solved in these dust evolution models are given in \cite{birnstiel2010} and \cite{stammler2022}. 

The original radial grid from the hydrodynamical simulations that had 672 cells (from 2.2 and 198\,au) was interpolated to 100 cells to speed up the dust evolution models. {With this resolution, we resolved the outer edge of the gap, where particles accumulate with about 26 radial cells, and the gap and inner disk were resolved with more radial cells because the grid is logarithmically spaced. A resolution test is present in Appendix~\ref{appendixA}.} The peak of the pressure maximum from the hydro models is about 60\,au. This peak should coincide with the peak of the ring observed at millimeter observations at 77\,au. In our models, we took the pressure maximum as the reference point, and we therefore scaled the results from the hydrodynamical simulation before performing the dust evolution models, such that the pressure maximum was at 77\,au. After this correction, the radial grid extended from 2.9 to 249\,au. 

The stellar parameters were the same as in \cite{bae2019}, that is, a 0.85\,$M_\odot$ stellar mass. The viscosity was taken as in the hydrodynamical simulations ($\alpha=10^{-3}$), which was also assumed for dust diffusion and for the vertical and radial turbulence of the dust \citep{pinilla2021}.

The fragmentation velocity of the particles was assumed to be 10\,m\,s$^{-1}$. This fragmentation velocity assumes that the outer layer of the grains is composed of water-ice, for which the sticking force is thought to be strong. Currently, there are no laboratory experiments that reproduce the collisions of water-ice dust particles at the low temperatures and pressures of those of protoplanetary disks \citep{blum2000, wada2009, gundlach2018, arakawa2023}. Laboratory experiments that decreased the temperature of dust collision experiments to $\sim$150-180\,K suggested that water-ice particles may not be as sticky as thought before and that the fragmentation velocity is lower \citep[$\sim 1\,$m\,s$^{-1}$, e.g.,][]{musiolik2019, steinpilz2019}. \cite{pinilla2021} demonstrated that when the fragmentation velocity of the particles is $1\,$m\,s$^{-1}$ in the entire disk, the disk needs to have low turbulence to be able to reproduce millimeter observations of typical disks. However, with low turbulence ($\alpha < 10^{-4}$), a Jupiter-mass planet is expected to create multiple types of substructures, which are not observed in the \PDS ~disk, including multiple rings, gaps, and asymmetries \citep[e.g.][]{ataiee2013, bae2018}. We therefore decided to work with the hydrodynamical results that assumed $\alpha=10^{-3}$, for which $v_f=10$\,m\,s$^{-1}$ works well to reproduce observations.

The dust density distribution was initially assumed {as the interstellar medium} to follow a power law as $n(a)\propto a^{-3.5}$ from a minimum grain size, which we varied between [$1\times10^{-6}$, $1\times10^{-5}$, $5\times10^{-5}$] cm. The maximum grain size in the initial dust size distribution was $1\times10^{-4}$cm. We therefore present results from three different models. {We note that in \cite{pinilla2016}, we investigated the effect of assuming a different power law for the initial condition of dust distribution on the dust trapping and filtration, and we found that the results are not affected by this assumption.}

These initial minimum grain sizes were selected to explore the dust filtration at the outer edge of the gap and investigate the size of dust particles that is required to  maintain the inner disk.  Because the grains were allowed to grow, the maximum grain size of the grid was 10\,cm, and we took 7 grid cells for each order of magnitude in mass \citep{stammler2022}. This means that the grain size grid had 148, 127, and 113 cells when $a_{\rm{initial}}=[1\times10^{-6}-10^{-4}]$\,cm, $a_{\rm{initial}}=[1\times10^{-5}-10^{-4}]$\,cm, and $a_{\rm{initial}}=[5\times10^{-5}-10^{-4}]$\,cm, respectively.  The initial dust-to-gas ratio was 1/100. The volume density of the particles ($\rho_s$) was set to {0.85g\,cm$^{-3}$ in order to be consistent with the grain composition and opacities assumed in the radiative transfer simulations}. The simulations ran from 0 to 10\,Myr. Under the hypothesis that the two planets required at least 1\,Myr to form in addition to the 1\,Myr of planet-disk interaction in the hydrodynamical models, the initial time of the dust evolution models should be seen as at least 2\,Myr of the disk age.

\subsection{Radiative transfer simulations}

To compare the results from the dust evolution models to previous and future ALMA observations and give predictions for different wavelengths, we performed radiative transfer calculations with \texttt{RADMC3D} \citep{dullemond2012}. The opacity of each grain size was calculated using \texttt{optool} \citep{dominik2021}, and we assumed that the grains were 
porous spheres with a vacuum volume fraction of 40\% and that they were composed of 10\% silicate, 20\% carbon, and 30\% water ice \citep{ricci2010}. {This composition gives a volume density of the dust particles of 0.85g\,cm$^{-3}$. \cite{stadler2022} and \cite{zormpas2022} demonstrated that with these opacities, the models of dust evolution and radiative transfer agree better with observations, specifically, with the millimeter fluxes and spectral indices \citep[in contrast to the DSHARP opacities,][]{birnstiel2018}.}

The only source of radiation is the central star, which was assumed to be a black-body with a temperature of 3972\,K. For our calculations, we assumed $1\times10^{7}$ photons and $5\times10^{6}$ scattering photons.

We calculated the total volume dust density of each grain size as

\begin{equation}
        \rho_d(R,\varphi,z, \mathrm{St}) = \frac{\Sigma_d(R, \mathrm{St})}{\sqrt{2\,\pi}\,h_d (R, \mathrm{St})}\,\exp \left( -\frac{z^2}{2\,h_{\mathrm{d}}^2(R, \mathrm{St})} \right)\,,
        \label{eq:volume_density}
\end{equation}

\noindent where  $z = r\,\cos(\theta)$ and $R=r\,\sin(\theta)$, with $\theta$ being a polar angle. $\Sigma_d$ is the dust surface density obtained from the dust evolution models. The radial grid was assumed as in the dust evolution models, and for the vertical and azimuthal grid, we assumed 32 and 64 cells, respectively.  The dust scale height $h_{\mathrm{d}}$ for each particle size was given by \citep{youdin2007, birnstiel2010}

\begin{equation}
        h_d(\mathrm{St})=h \times \rm{min} \left( 1,\sqrt{\frac{\alpha}{\mathrm{min}(\rm{St},1/2)(1+\rm{St}^2)}}\,\right),
        \label{eq:dust_scaleheight}
\end{equation}

\noindent where $\alpha=10^{-3}$ as in the hydrodynamical and dust evolution models, and \rm{St} is the Stokes number of the dust particles calculated at the midplane, which is

\begin{equation}
    \rm{St}=\frac{\text{$a$} \rho_s}{\Sigma_g}\frac{\pi}{2},
    \label{eq:Stokes_number}
\end{equation}

\noindent with $a$ as the grain size, and $\Sigma_g$ the gas surface density. A similar procedure was used for example by \cite{pohl2017} and \cite{pinilla2021}.

\section{Results} \label{sect:results}

\subsection{Dust density distribution}
Figure~\ref{fig:dust_evolution_models} shows the dust density distribution after 1\,Myr (top panes) and 10\,Myr (bottom panels) of evolution as a function of grain size and distance from the star. The y-axis of each column is different because it depends on the initial grain size that is assumed in the models (given in the title of each column). The plots include the St=1 (solid white line), which is proportional to the gas surface density taken from the hydrodynamical simulations. 

The left column of Fig.~\ref{fig:dust_evolution_models}
corresponds to models where $a_{\rm{initial}}=[5\times10^{-5}-10^{-4}]$\,cm, which is our model with the largest minimum grain size. As expected, the filtration of dust particles at the outer edge of the gap created by the planets is very efficient in this case. At 1\,Myr of evolution, the dust grains that were initially within the gap have drifted toward the star, and after 10\,Myr of evolution, very little dust is left in the inner disk. 

This efficient filtration can be understood in terms of the radial dust velocity of particles, which is given by

\begin{equation}
    \varv_{\mathrm{r,dust}}=\frac{\varv_{\mathrm{r,gas}}}{1+\textrm{St}^2}+\frac{1}{\textrm{St}^{-1}+\textrm{St}} \frac{\partial_r P}{\rho \Omega},
        \label{eq_vdust} 
\end{equation}

\noindent where $\varv_{\mathrm{r,gas}}$ is the radial gas velocity, $P$ is the gas pressure, $\rho$ is the gas volume density, and $\Omega$ is the Keplerian frequency.

\begin{figure}
\centering 
     \includegraphics[width=9cm]{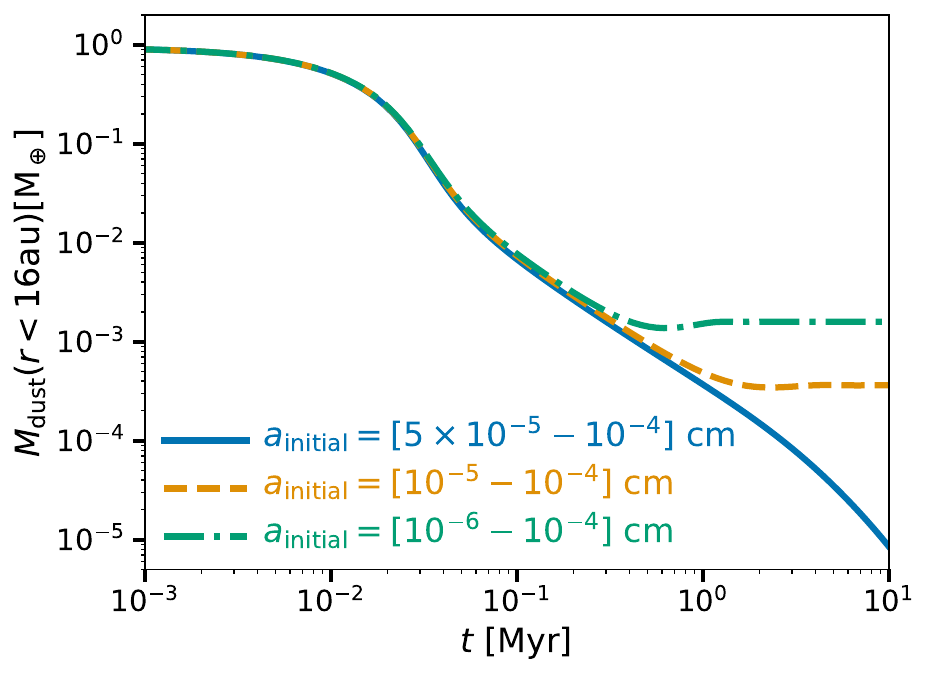}
\caption{Total dust disk mass within 16\,au as a function of time for the models of Fig.~\ref{fig:dust_evolution_models}}
\label{fig:Mdust_time}
\end{figure}

\begin{figure}
\centering 
     \includegraphics[width=9cm]{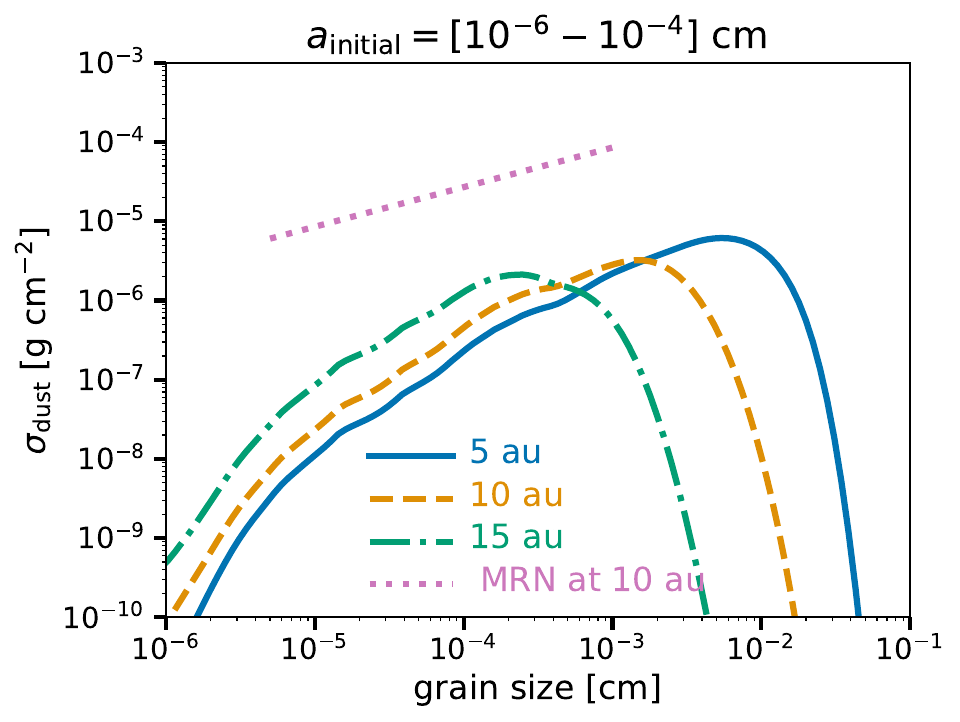}
\caption{Grain size distribution at radial distances of 5, 10, and 15\,au (this means inside the orbit of \PDS b) for the models where $a_{\rm{initial}}=[1\times10^{-6}-10^{-4}]$\,cm. The dotted line shows the MNR distribution at 10\,au for comparison.}
\label{fig:sigma_dust_case3}
\end{figure}

Only when $\varv_{\mathrm{r,dust}}$ is positive are particles trapped. This implies that particles larger than critical size $a_{\rm{critical}}$ are filtered out \cite{pinilla2012}, with

\begin{equation}
        a_{\rm{critical}}\gtrsim-\frac{2 \varv_{\mathrm{r,gas}}\rho\Omega\Sigma_g}{\partial_r P \pi \rho_s}.
  \label{a_critical}
\end{equation}

Because all the terms in Eq.~\ref{a_critical} are quantities related to the gas, except for $\rho_s$ 
(which is the volume density of the dust particles set to 1.6\,g\,cm$^{-2}$), $a_{\rm{critical}}$ can be calculated given the gas density distribution from the hydrodynamical simulations. When we assume hydrostatic equilibrium of a locally isothermal disk and viscous accretion to calculate $\rho$ and $\varv_{\mathrm{r,gas}}$, respectively, $a_{\rm{critical}}$ at the outer edge of the gap is $\sim0.1\,\mu$m. For this reason,  when $a_{\rm{initial}}=[5\times10^{-5}-10^{-4}]$\,cm, the dust filtering at the outer edge of the gap is very effective, which leads to a high depletion of dust in the inner disk. In a few words,  the  dust of the inner disk disconnects from the outer disk, and the dust that was initially inside the gap drifts toward the star, while in the outer disk, it drifts toward the pressure maximum at the outer edge of the gap.

The values of $a_{\rm{critical}}$ at the outer edge of the gap do not change significantly when considering the gas radial velocity directly from the hydrodynamical simulations. However, we did not perform dust evolution models with these gas velocities due to the large fluctuations near the planet \citep[see for example Fig. 3 in][]{joanna2019}, which can lead to artificial dust accumulations when gas and dust evolution are not modeled simultaneously.

\begin{figure*}
\centering 
     \includegraphics[width=18cm]{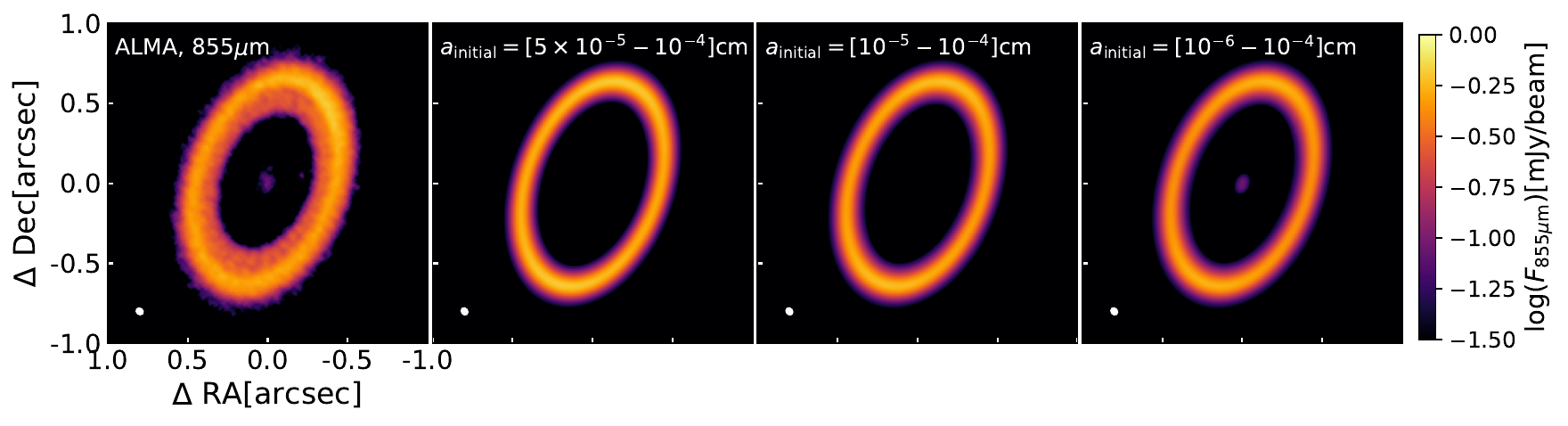}
\caption{Comparison between models and ALMA observations of \PDS. Left panel: ALMA dust continuum emission at 855\,$\mu$m from \cite{benisty2021} vs. synthetic images at the same wavelength taking the models of Fig~\ref{fig:dust_evolution_models} after 1\,Myr of evolution. The beam for the synthetic images is assumed to be as in the observations, i.e., $\sim$0.046''$\times$0.36'' with a position angle of 51.5$^\circ$.}
\label{fig:comparison_ALMA}
\end{figure*}

The result of efficient filtration is seen in Fig.~\ref{fig:Mdust_time}, which corresponds to the total dust disk mass within 16\,au as a function of time for the models of Fig.~\ref{fig:dust_evolution_models}. The reference location of 16\,au is taken below to compare to the results from \cite{portilla2023}. The total dust mass within this inner region at the initial time is the same for all the three models ($\sim0.9\,M_\oplus$). In the model where $a_{\rm{initial}}=[5\times10^{-5}-10^{-4}]$\,cm, this mass is reduced to {$3.3\times10^{-4}\,M_\oplus$ (or 0.025\,$M_{\rm{Moon}}$) after 1\,Myr of evolution}, and it continues to decrease, reaching values of {$8.4\times10^{-6}\,M_\oplus$}after 10\,Myr.

The results of these models motivated us to decrease the size of the initial dust particles to as low as the critical grain size (Eq.~\ref{a_critical}) of 0.1\,$\mu$m or even lower. The middle and right columns of Fig.~\ref{fig:dust_evolution_models} show the dust density distribution with $a_{\rm{initial}}=[10^{-5}-10^{-4}]$\,cm and $a_{\rm{initial}}=[10^{-6}-10^{-4}]$\,cm, respectively. These results show that the dust density distribution in the inner disk increases in these two cases and that the dust filtration at the outer edge of the gap is not total.  The  dust mass within 16\,au for the $a_{\rm{initial}}=[10^{-5}-10^{-4}]$\,cm case decreases to {$4.6\times10^{-4}\,M_\oplus$} after 1\,Myr of dust evolution, similar to the case of  $a_{\rm{initial}}=[5\times10^{-5}-10^{-4}]$\,cm, but after this time, the dust mass does not significantly decrease over the next million years (Fig.~\ref{fig:Mdust_time}). This is because the inner disk is continuously replenished by the very  small grains ($\sim$0.1$\mu$m) that can flow through the gap. The replenishment of the inner disk from the outer disk dust trap  is more efficient when the initial dust size distribution includes smaller grains, $a_{\rm{initial}}=[10^{-6}-10^{-4}]$\,cm, as grains with lower Stokes number can be more efficiently dragged and diffused by the gas. This is reflected in the total amount of dust that remains in the inner disk in these simulations (Fig.~\ref{fig:Mdust_time}). For this case, the total amount of dust mass is {$\sim1.6\times10^{-3}\,M_\oplus$} after 1\,Myr of evolution, which remains constant until the end of the simulations (10\,Myr). This amount of dust is much lower than the dust inferred by \cite{benisty2021} for the inner disk of 0.08–0.36\,$M_{\oplus}$ for a minimum and a maximum grain size of 0.05\,$\mu$m and 10\,$\mu$m, respectively, and the DSHARP opacities \citep{birnstiel2018}. In our models, the maximum grain size in the inner disk varied between $a_{\rm{max}}\sim$10-100\,$\mu$m depending on the distance from the star, as shown in Fig.~\ref{fig:sigma_dust_case3}. Therefore, our models have  grains of higher opacity at millimeter wavelengths than those considered in \cite{benisty2021}, especially when comparing the DSHARP opacities with those used in our models from \cite{ricci2010} \citep[see for example Fig.~A.3 in ][]{stadler2022}. Even with lower dust masses, our models can therefore reproduce the millimeter emission of the inner disk observed with ALMA. It is worthwhile to note that the obtained grain distribution in the inner disk (Fig.~\ref{fig:sigma_dust_case3}) is  very different from  the  size distribution of interstellar grains from \cite[MRN distribution][]{mathis1977} of $n(a)\propto a^{-3.5}$, which was assumed in the calculations by \cite{benisty2021}, and this is  plotted in Fig.~\ref{fig:sigma_dust_case3} for comparison. {Most of the recent dust models for the interstellar medium find a different size distributions than the MRN \citep[e.g.][]{koehler2015, jones2017, hensley2023}, which can also affect the inferred dust mass from the inner disk presented in \cite{benisty2021}.}

\subsection{Comparison with ALMA observations}

Figure~\ref{fig:comparison_ALMA} shows the comparison of models and current ALMA observations of \PDS. For this comparison, synthetic images were created at the same wavelength (855\,$\mu$m) as the observations presented in \cite{benisty2021}, assuming the dust density distribution from the models shown in Fig.~\ref{fig:dust_evolution_models} after 1\,Myr of evolution. The beam for the synthetic images was assumed to be as in the ALMA observations, that is, $\sim$0.046''$\times$0.36'', with a position angle of 51.5$^\circ$.

\begin{figure*}
\centering
    \tabcolsep=0.05cm 
    \begin{tabular}{cc} 
     \includegraphics[width=9cm]{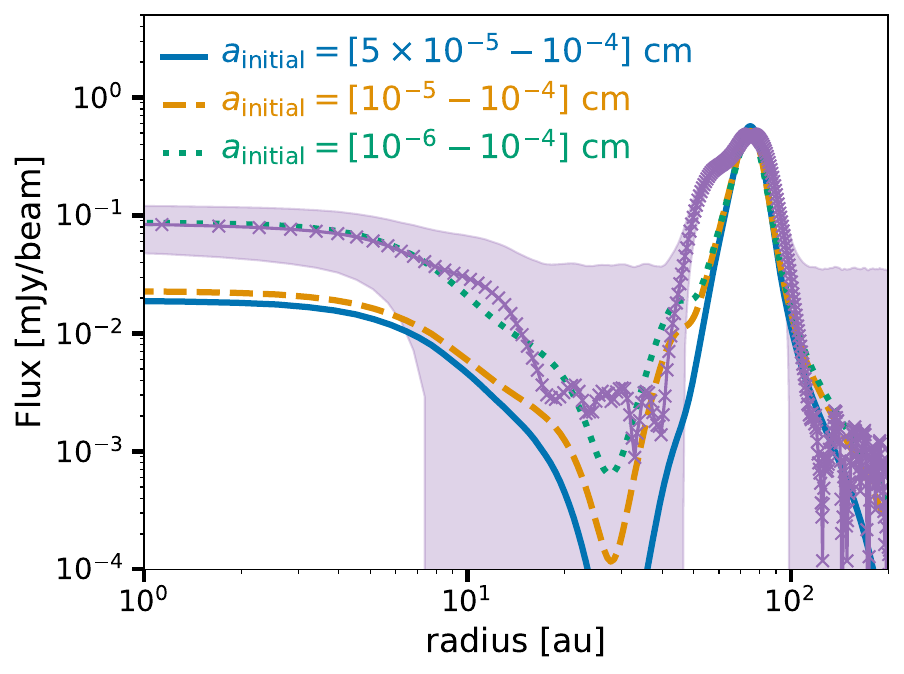}&
     \includegraphics[width=9cm]{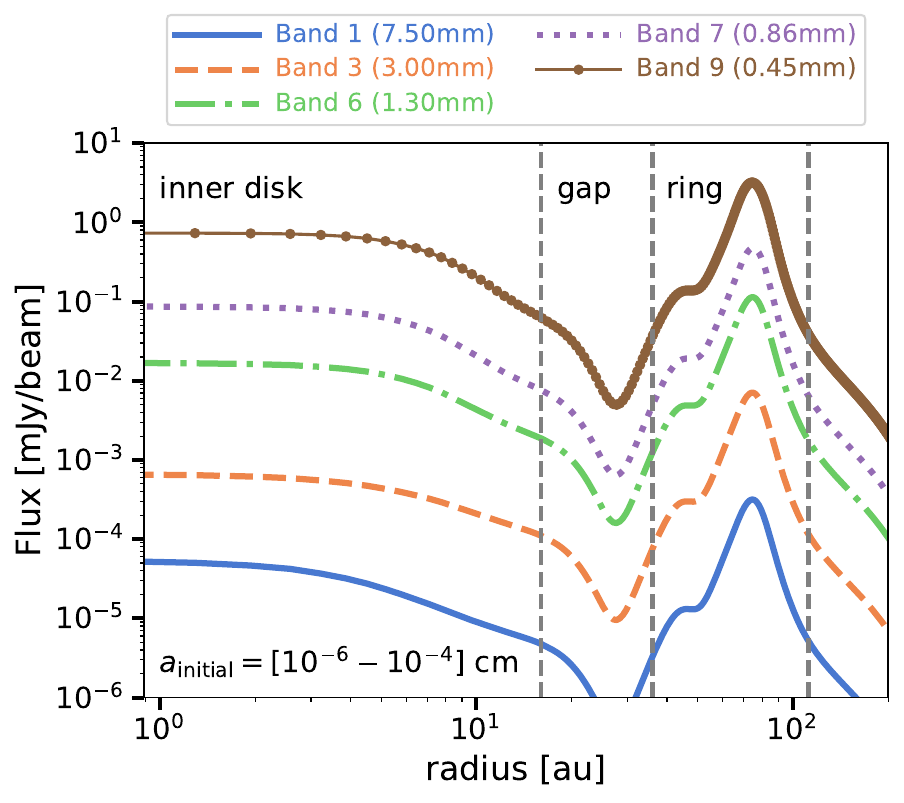}
     \end{tabular}
\caption{Models and observational predictions for \PDS disk. Left:Comparison of the azimuthally averaged radial intensity profiles of the deprojected images of the continuum from ALMA Band 7 observations (crosses) and the synthetic images from Fig.~\ref{fig:comparison_ALMA}. The shaded area is the standard deviation of each elliptical bin divided by the square root of the number of beams spanning the full azimuthal angle at each radial bin. Right: Predictions for several ALMA bands for the case where $a_{\rm{initial}}=[10^{-6}-10^{-4}]$\,cm, assuming that all the images have the same resolution as the observations in Band 7 (Fig.~\ref{fig:comparison_ALMA}).}
\label{fig:radial_profiles_images}
\end{figure*}

The synthetic images in Fig~\ref{fig:comparison_ALMA} demonstrate that the dust evolution models assuming $a_{\rm{initial}}=[5\times10^{-5}-10^{-4}]$\,cm or $a_{\rm{initial}}=[10^{-5}-10^{-4}]$\,cm do not reproduce the emission of the inner disk that is observed. This is also shown in the left panel of Fig~\ref{fig:radial_profiles_images}, which corresponds to the comparison of the azimuthally averaged radial intensity profiles of the deprojected images from the observations and the models. 

The model in which $a_{\rm{initial}}=[10^{-6}-10^{-4}]$\,cm can reproduce the emission of the inner disk well even when the total dust mass within this region is very low ($\sim2\times10^{-3}\,M_\oplus$). It is worth mentioning that the models with small grains also predict an inner shoulder just within the main ring. This emission is located near the location of the outer planet, in which a small bump is created in between the two gaps. There is a similar shape of emission in the actual observations \citep[Band 7;][]{benisty2021} with an inner shoulder within the bright ring (Fig.~\ref{fig:comparison_ALMA} and Fig~\ref{fig:radial_profiles_images}), but the models underpredict the flux of this structure because the grain size of the dust particles is small at that location. Nonetheless, it is interesting to see that the emission of this shoulder also increases as the minimum grain size in the simulations is smaller and when the filtration is less efficient (left panel in Fig~\ref{fig:radial_profiles_images}). A higher angular resolution is required to understand the  properties of this emission and test whether it is a well-separated ring, for example. 

The right panel of Fig~\ref{fig:radial_profiles_images} shows the averaged radial intensity profiles of deprojected images obtained at different ALMA bands, showing that the emission decreases at longer wavelengths in the model with $a_{\rm{initial}}=[10^{-6}-10^{-4}]$\,cm and assuming the same angular resolution as for the observations from \cite{benisty2021} (left panel in Fig.~\ref{fig:comparison_ALMA}). The detection of the inner disk would depend on the sensitivity of the observations to detect it. Already in Band 3 (3\,mm), the flux of the inner disk is a few micron-Jy per beam. 

\begin{figure}
    \centering
\includegraphics[width=9cm]{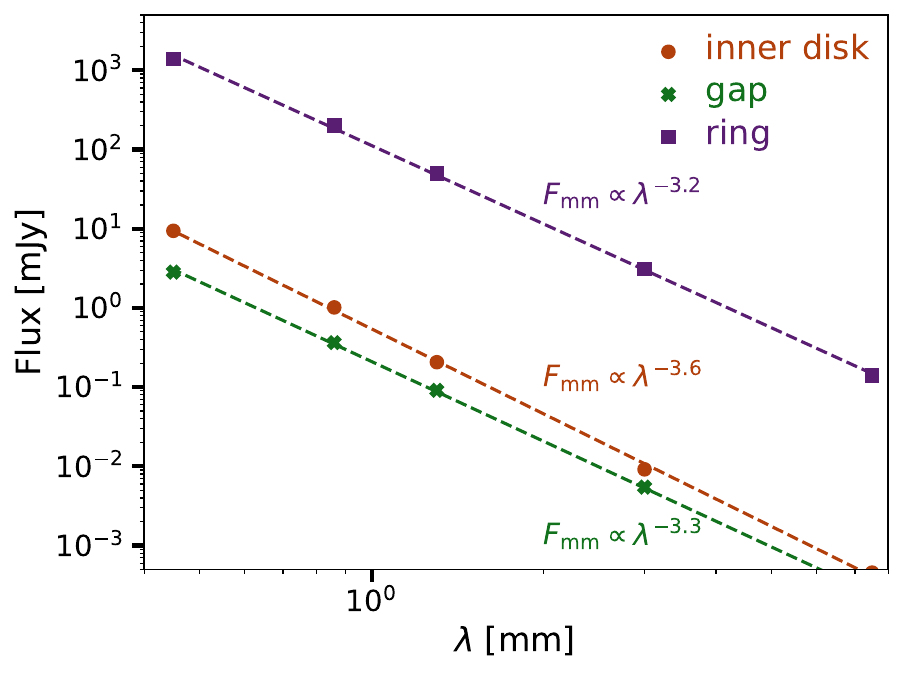}
    \caption{Total millimeter flux as a function of  wavelength from the same ALMA bands as in the right panel of Fig.~\ref{fig:radial_profiles_images} (case of $a_{\rm{initial}}=[10^{-6}-10^{-4}]$\,cm). The fluxes are calculated in three different regions (as denoted in the right panel of Fig.~\ref{fig:comparison_ALMA}), the inner disk (r$<$16\,au), the gap ($16\,\rm{au}<r<$36\,au), and the ring ($36\,\rm{au}<r<$112\,au). The dashed line shows the best power-law fit to the data and the spectral index for each region.}
    \label{fig:spectral_index}
\end{figure}

Figure~\ref{fig:spectral_index} shows the total millimeter flux as a function of  wavelength from the same ALMA bands as in the right panel of Fig.~\ref{fig:comparison_ALMA}. The fluxes were calculated in three different regions (as denoted in the right panel of {Fig.~\ref{fig:radial_profiles_images}}), the inner disk ($r<$16\,au), the gap ($16\,\rm{au}$ $<r<$36\,au), and the ring ($36\,\rm{au}<$$ r<$112\,au). Using \texttt{curvefit} within \texttt{scipy} \citep{virtanen2020}, we fit $F_\lambda \propto \lambda^{-\alpha_{\rm{mm}}}$ to find the spectral index $\alpha_{\rm{mm}}$ of each region. The spectral index of the ring has values of $\alpha_{\rm{mm}}=3.2$. Because the total flux is dominated by the ring emission, the spectral index integrated over the entire disk is also 3.2, which is similar to the value within the gap ($3.3$).

In this model, the spectral index is slightly higher in the inner disk, with a value of  $\alpha_{\rm{mm}}=3.6$. Hence, these models  suggest that the spectral index that can be obtained from  observations using optically thin wavelengths will show a high spectral index of $\sim3.2$ with a slightly increased inner disk. This value of the spectral index is high compared to the averaged values found in protoplanetary disks \citep[e.g.,][]{tazzari2021}, but for a set of transition disks, \cite{pinilla2014} showed that the observed spectral index increased with the size of the cavity and that overall, transition disks have a higher spectral index than the disks without cavities. Specifically, the linear relation between the spectral index and the cavity size (defined as the peak observed  in millimeter emission) is $\alpha_{\rm{mm}}=0.011\times R_{\rm{cavity}}+2.36$. For a cavity size of $\sim74$\,au \citep{keppler2019}, the expected $\alpha_{\rm{mm}}$ is 3.2, as found in our models, suggesting that the spatially integrated spectral index of \PDS~ agrees with observations of other transition disks.

These values, however, do not match the spectral index of 2.7 obtained from the total flux at 855\,$\mu$m \citep[$\sim$176\,mJy,][]{keppler2019} and at 1.3,mm \citep[$\sim$57\,mJy,][]{facchini2021}, which could be due to the high optical depth of the observations, especially at 855\,$\mu$m. Observations at multiple wavelengths where the emission is optically thin are needed to test current models and  the results of Fig.~\ref{fig:spectral_index}.

\section{Discussion} \label{sect:discussion}

\subsection{Effect of the gas disk mass and turbulence}

The filtration of dust particles at the outer edge of a gap depends on how well the particles are coupled to the gas and on the level of dust diffusion driven by the disk turbulence.

The coupling of dust particles to the gas is quantified by the Stokes number given in Eq.~\ref{eq:Stokes_number}. Particles with St$\ll$1 are well coupled to the gas and can be dragged along with it, while St$\sim$1 are well trapped and hence filtered out. {There is an uncertainty related to the bulk density of the dust particles, and hence an uncertainty related to the dust composition and porosity. In this work, we tested dust evolution models in which the bulk density was assumed to be 1.6\,g\,cm$^{-3}$, similar to the value obtained when the DSHARP opacities are assumed. This change leads to Stokes numbers that are higher by a factor of $\sim$ 2 for the same grain size and gas surface density, which does not affect the results presented in this work significantly. In our models,} the main uncertainty in the Stokes number is the gas surface density, which directly depends on the gas disk mass. This still remains unknown from observations of most disks and, \PDS~ is no exception.

\cite{portilla2023} performed detailed thermochemical models of the ALMA Band 6 observations, in particular, to model the distribution of three CO isotopologs, in order to constrain the gas distribution of \PDS. Their analysis suggested a gas mass of  $3.2\times10^{-3}$\Mjup within the dust-depleted gap (between 16 and 41\,au). The gas mass within the same region in our models is $3.0\times10^{-3}$\Mjup~   , which agrees well with the values from \cite{portilla2023}. 

The total disk mass (inside 130\,au) from \cite{portilla2023} is $3.0\times10^{-4}\,M_\odot$, while in our models, it is $9.0\times10^{-4}\,M_\odot$. An assumed lower disk mass would imply that the Stokes number is higher for a given grain size, and as result, less filtration is expected. In other words, a less massive disk would imply that the minimum grain size would need to be lower than the best-case scenario explored in this work (i.e.,$<1\times10^{-6}$\,cm). {Nano-size particles like this have been suggested to be present from observations of some protoplanetary disks \citep[][]{habart2021, Kokoulina2021}.}

\begin{figure}
\centering 
     \includegraphics[width=9cm]{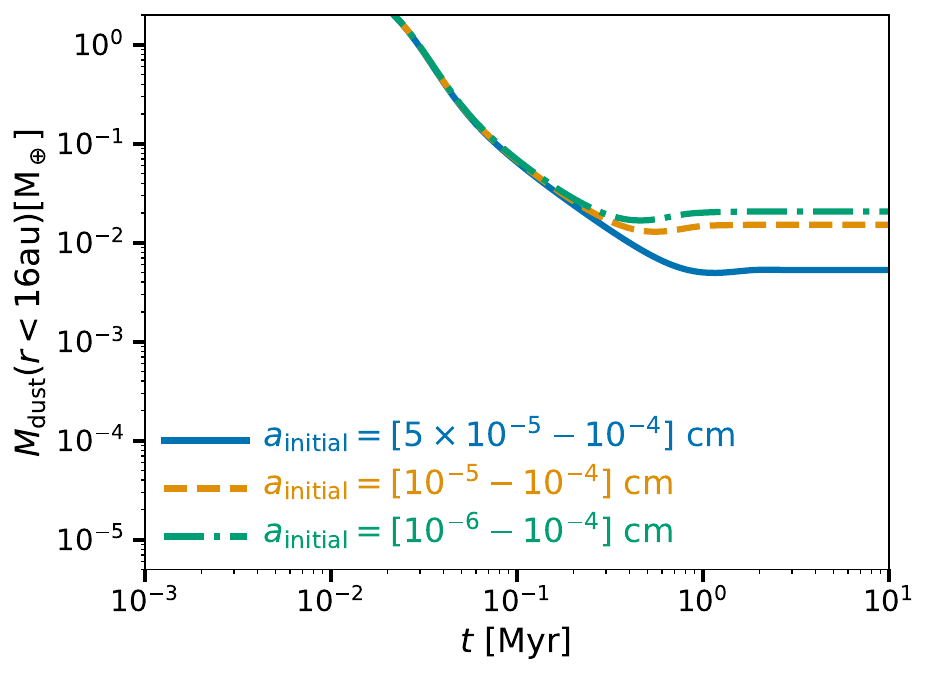}
\caption{As Fig~\ref{fig:Mdust_time}, but with a gas disk mass that is ten times higher}
\label{fig:Mdust_10Mgas}
\end{figure}

It is still possible that the gas disk mass obtained from CO observations is undestimated because of different chemical and physical processes that can convert most CO into other refractory molecules or because CO is sequestered into icy bodies \citep{schwarz2018, krijt2020}. As a test, we performed three dust evolution models in which the gas disk mass was increased by a factor of ten and assumed the same initial grain size as in the models presented in Sect.~\ref{sect:results}. The summary of these simulations is presented in Fig.~\ref{fig:Mdust_10Mgas}, and it corresponds to the evolution of the dust mass for each case (as Fig.~\ref{fig:Mdust_time}). These results show that when the disk mass is higher, the filtration is lower for a given initial particle size distribution. For instance, for the models with $a_{\rm{initial}}=[5\times10^{-5}-10^{-4}]$\,cm, the dust mass from 1 to 10\,Myr is $4\times10^{-3}\,M_\oplus$ (similar to the dust mass in the models of Sect.~\ref{sect:results}, but for $a_{\rm{initial}}=[10^{-6}-10^{-4}]$\,cm) and does not continuously decrease as in the case of Fig.~\ref{fig:Mdust_time}. This highlights the importance of obtaining better constraints on the gas disk mass, which can have a large impact on the dust filtration effect and the dust density distribution in the inner disk of \PDS. 

Recently, \cite{trapman2022} suggested a new method for measuring gas disk masses from observing N$_2$H$^{+}$ and CO isotopologs, which was tested against HD observations of several disks. These observations have not been performed for \PDS so far, but it would be interesting to test this method in future observations and compare it with the disk masses from \cite{portilla2023}, which agree with the values used in this work.

The level of turbulence can also change the effect of filtration, with higher turbulence implying less filtration. However, multiple models and observations of disks suggest that turbulence is rather low \citep[$\alpha\lesssim10^{-3}-10^{-4}$, see e.g. the recent review by][]{rosotti2023}. For transition disks in particular, a higher viscosity would imply a less empty cavity or no cavity at all, regardless of the mass of the planets that are embedded in the disk \citep{ovelar2016}. On the other hand, a lower viscosity for \PDS~disk would imply more filtration, but also multiple rings and gaps created by the planets that are embedded in this disk \citep[e.g.,][]{dong2017b, bae2018}. As a result, a viscosity of $\alpha\sim10^{-3}$, as we used here, seems appropriate for \PDS. The intrinsic nature of $\alpha-$viscosity is complex in any case. It depends on how the disks are ionized and coupled to the magnetic fields \citep[e.g.,][]{ lesur2023}, which can lead to spatial variations of $\alpha$ depending on different enviromental, stellar, and disk properties \citep{delage2021}.

\subsection{Inner emission and comparison with JWST observations}

The JWST/MIRI observations of \PDS~ revealed the presence of water in its inner disk \citep[$<1\,$au,][]{perotti2023}. Specifically, the H$_2$O spectrum is best fit with a slab model of gas at 600\,K inside 0.05\,au and a column density of $N = 1.4\times10^{18}$\,cm$^{-2}$. Small grains are of great importance to explain this water emission for two reasons: First, small grains can help to shield  ultraviolet light and prevent H$_2$O photodissociation \citep{heays2017}, and second, they can be the particles that can travel across the gap and enrich the inner disk, as demonstrated in this work.

We calculated the total dust mass that potentially reaches the snowline in \PDS~ by taking the dust-loss rates ($\dot M_{\rm{dust}}$)  at the inner edge of the simulations in Sect.~\ref{sect:models}, and integrating over either 0-1\,Myr, where the vast majority of the accreted dust comes from the initial material that was inside the gap location, or over 1-3.5\,Myr, where most of the material comes from the potential replenishment from the outer to the inner disk. We did not integrate over the entire 10\,Myr of the dust evolution models  because the age of \PDS~ is approximately 5.4\,Myr, and as discussed before, the initial time in the dust evolution models should be seen as at least 2\,Myr of the disk age.

For the three different values of the initial grain size of the models in Fig.~\ref{fig:dust_evolution_models}, the total dust mass lost to the inner bounday ($M_{\rm{dust, inner}}$) is approximately {0.9\,$M_\oplus$} between 0-1\,Myr for all the three models, and from 1 to 3.5\,Myr, it varies between  {$2.7\times10^{-4}\,M_\oplus$, $4.5\times10^{-4}\,M_\oplus$ and $4.3\times10^{-3}\,M_\oplus$} for $a_{\rm{initial}}=[5\times10^{-5}-10^{-4}]$\,cm, $a_{\rm{initial}}=[1\times10^{-5}-10^{-4}]$\,cm, and $a_{\rm{initial}}=[1\times10^{-6}-10^{-4}]$\,cm, respectively. Taking the latter case, which reproduces the inner emission observed with ALMA, and assuming that (1) 30\% of that dust is water ice ($f_{\rm{ice}}$, as we assumed for the dust opacities), (2) {a  percentage} of this dust is likely to grow to planetesimals near the snowline (\citep[e.g.,][]{carrera2017, joanna2017}, we denote this efficiency as $\epsilon_{\rm{planetesimals}}$),  (3) that at least half of the remaining dust particles  that reach the snowline is again moved outward via vapor diffusion \cite{kalyaan2019}, and (4) that at least half of this dust was accreted by the two planets and/or their circumplanetary disks \citep{joanna2018}, we  calculate the column density of water that is expected within the snowline (i.e., $r_{\rm{snow}}$, which we take at 1\,au), assuming an optically thin slab, as

\begin{equation}
    N_{\rm{H}_2\rm{O}}= \frac{f_{\rm{ice}} \times M_{\rm{dust, inner}} \times (1-\epsilon_{\rm{planetesimals}})}{8\times\pi r_{\rm{snow}}^2\times m_{\rm{H}_2\rm{O}}\times \cos i}
\end{equation}

\noindent where $m_{\rm{H_2O}}$ is the mass of a water molecule, that is, 18\,g/mol, and $i$ is the disk inclination taken to be 51.7$^\circ$. Assuming that $\epsilon_{\rm{planetesimals}}$ is {10\% or 90\%}, we obtain that $N_{\rm{H_2O}}$ is $\sim 6.7\times10^{19}$\,cm$^{-2}$ when $\epsilon_{\rm{planetesimals}}$ is 10\% and  $\sim 7.5\times10^{18}$\,cm$^{-2}$ when $\epsilon_{\rm{planetesimals}}$ is 90\%. There are several assumptions in this calculation. In particular, we neglected any chemical effect that can change the water abundance, but this was done just to illustrate that the water column density observed with JWST might solely originate from drifting particles that come from the outer disk, which need to be small enough in our models ($<$0.1\,$\mu$m) to pass through the gap formed by the two giant planets in this system.

\section{Conclusions} \label{sect:conclusions}
\PDS~ remains the best laboratory for studying the influence of the formation of giant planets on the disk structure and the potential formation of terrestrial planets within the system. One of the intriguing properties of \PDS~ is its inner disk, which was revealed by the spatially resolved emission observed with ALMA at different wavelengths. Its survival is puzzling because dust evolution models suggest that complete filtering of dust particles by the gap created by the planets would stop the dust replenishment from the outer to the inner disk. We confirmed this result by performing advanced dust evolution models, in which the minumum grain size in the models was larger than 0.1$\mu$m. As a consequence, the initial dust that was within the orbit of the planets eventually migrated inward, and the inner disk was depleted within the first million years of the dust evolution. 

To maintain an inner disk, the minimum grain size in the models needs to be lower than 0.1$\mu$m. These results {mainly} depend on the assumption taken for the gas disk mass, and we performed models with values that agreed with recent thermochemical models, which were compared with ALMA observations \citep{portilla2023}. Only when the grains are that small are they diffused and dragged along with the gas throughout the gap of the planets. These small grains grow and drift in the inner disk, but the constant reservoir of dust particles that are trapped in the outer edge of the gap and that fragment continuously helps to preserve the inner disk on million-year timescales. Our flux predictions at millimeter wavelength of these models agree with current ALMA observations. {Other possible uncertainties originate from the planet masses. If the planet masses are lower than the mass of Jupiter, the filtration effect is not as efficient as presented in this work. However, observations and models of the \PDS~disk robustly support that \PDS b and \PDS c are more massive than Jupiter.}

It is possible that the water emission that was recently observed with the JWST from the inner disk of \PDS~ originates from these small grains that cross the gap and then drift toward the inner disk, where they reach the snowline. This idea can be a mirror of our Solar System, in which the formation of Jupiter and Saturn could have blocked the majority of pebbles in the outer disk, but where water ice could still have been carried by the very small grains in the parental disk of the Solar System.  Future observations that better constrain the gas disk mass and the level of turbulence in \PDS~ will help us to test our models.

\begin{acknowledgements}
{We are very thankful to the referee, whose comments helped to clarify different aspects of this paper.} This paper makes use of the following ALMA data: ADS/JAO.ALMA\#2018.A.00030.S.  ALMA is a partnership of ESO (representing its member states), NSF (USA) and NINS (Japan), together with NRC (Canada), MOST and ASIAA (Taiwan), and KASI (Republic of Korea), in cooperation with the Republic of Chile. The Joint ALMA Observatory is operated by ESO, AUI/NRAO and NAOJ. This project has received funding from the UK Research and Innovation (UKRI) under the UK government’s Horizon Europe funding guarantee from ERC (under grant agreement No 101076489). This project has received funding from the European Research Council (ERC) under the European Union's Horizon 2020 research and innovation programme (PROTOPLANETS, grant agreement No.~101002188). S.F. is funded by the European Union (ERC, UNVEIL, 101076613). Views and opinions expressed are however those of the author(s) only and do not necessarily reflect those of the European Union or the European Research Council. Neither the European Union nor the granting authority can be held responsible for them. S.F. acknowledges financial contribution from PRIN-MUR 2022YP5ACE.
\end{acknowledgements}

\bibliographystyle{aa} %
\bibliography{ref}

\begin{thebibliography}{78}
\expandafter\ifx\csname natexlab\endcsname\relax\def\natexlab#1{#1}\fi

\bibitem[{{Aoyama} \& {Ikoma}(2019)}]{aoyama2019}
{Aoyama}, Y. \& {Ikoma}, M. 2019, \apjl, 885, L29

\bibitem[{{Arakawa} {et~al.}(2023){Arakawa}, {Tanaka}, {Kokubo}, {Nishiura}, \&
  {Furuichi}}]{arakawa2023}
{Arakawa}, S., {Tanaka}, H., {Kokubo}, E., {Nishiura}, D., \& {Furuichi}, M.
  2023, \aap, 670, L21

\bibitem[{{Asensio-Torres} {et~al.}(2021){Asensio-Torres}, {Henning},
  {Cantalloube}, {Pinilla}, {Mesa}, {Garufi}, {Jorquera}, {Gratton}, {Chauvin},
  {Szul{\'a}gyi}, {van Boekel}, {Dong}, {Marleau}, {Benisty}, {Villenave},
  {Bergez-Casalou}, {Desgrange}, {Janson}, {Keppler}, {Langlois}, {M{\'e}nard},
  {Rickman}, {Stolker}, {Feldt}, {Fusco}, {Gluck}, {Pavlov}, \&
  {Ramos}}]{asensio2021}
{Asensio-Torres}, R., {Henning}, T., {Cantalloube}, F., {et~al.} 2021, \aap,
  652, A101

\bibitem[{{Ataiee} {et~al.}(2013){Ataiee}, {Pinilla}, {Zsom}, {Dullemond},
  {Dominik}, \& {Ghanbari}}]{ataiee2013}
{Ataiee}, S., {Pinilla}, P., {Zsom}, A., {et~al.} 2013, \aap, 553, L3

\bibitem[{{Bae} {et~al.}(2018){Bae}, {Pinilla}, \& {Birnstiel}}]{bae2018}
{Bae}, J., {Pinilla}, P., \& {Birnstiel}, T. 2018, \apjl, 864, L26

\bibitem[{{Bae} {et~al.}(2019){Bae}, {Zhu}, {Baruteau}, {Benisty}, {Dullemond},
  {Facchini}, {Isella}, {Keppler}, {P{\'e}rez}, \& {Teague}}]{bae2019}
{Bae}, J., {Zhu}, Z., {Baruteau}, C., {et~al.} 2019, \apjl, 884, L41

\bibitem[{{Balsalobre-Ruza} {et~al.}(2023){Balsalobre-Ruza}, {de
  Gregorio-Monsalvo}, {Lillo-Box}, {Hu{\'e}lamo}, {Ribas}, {Benisty}, {Bae},
  {Facchini}, \& {Teague}}]{balsalobre2023}
{Balsalobre-Ruza}, O., {de Gregorio-Monsalvo}, I., {Lillo-Box}, J., {et~al.}
  2023, \aap, 675, A172

\bibitem[{{Banzatti} {et~al.}(2020){Banzatti}, {Pascucci}, {Bosman}, {Pinilla},
  {Salyk}, {Herczeg}, {Pontoppidan}, {Vazquez}, {Watkins}, {Krijt}, {Hendler},
  \& {Long}}]{banzatti2020}
{Banzatti}, A., {Pascucci}, I., {Bosman}, A.~D., {et~al.} 2020, \apj, 903, 124

\bibitem[{{Banzatti} {et~al.}(2023){Banzatti}, {Pontoppidan}, {Carr},
  {Jellison}, {Pascucci}, {Najita}, {Munoz-Romero}, {Oberg}, {Kalyaan},
  {Pinilla}, {Krijt}, {Long}, {Lambrechts}, {Rosotti}, {Herczeg}, {Salyk},
  {Zhang}, {Bergin}, {Ballering}, {Meyer}, {Bruderer}, \& {the JDISCS
  collaboration}}]{banzatti2023}
{Banzatti}, A., {Pontoppidan}, K.~M., {Carr}, J., {et~al.} 2023, arXiv
  e-prints, arXiv:2307.03846

\bibitem[{{Baruteau} {et~al.}(2019){Baruteau}, {Barraza}, {P{\'e}rez},
  {Casassus}, {Dong}, {Lyra}, {Marino}, {Christiaens}, {Zhu}, {Carmona},
  {Debras}, \& {Alarcon}}]{Baruteau2019}
{Baruteau}, C., {Barraza}, M., {P{\'e}rez}, S., {et~al.} 2019, \mnras, 486, 304

\bibitem[{{Benisty} {et~al.}(2021){Benisty}, {Bae}, {Facchini}, {Keppler},
  {Teague}, {Isella}, {Kurtovic}, {P{\'e}rez}, {Sierra}, {Andrews},
  {Carpenter}, {Czekala}, {Dominik}, {Henning}, {Menard}, {Pinilla}, \&
  {Zurlo}}]{benisty2021}
{Benisty}, M., {Bae}, J., {Facchini}, S., {et~al.} 2021, \apjl, 916, L2

\bibitem[{{Benisty} {et~al.}(2022){Benisty}, {Dominik}, {Follette}, {Garufi},
  {Ginski}, {Hashimoto}, {Keppler}, {Kley}, \& {Monnier}}]{benisty2022}
{Benisty}, M., {Dominik}, C., {Follette}, K., {et~al.} 2022, arXiv e-prints,
  arXiv:2203.09991

\bibitem[{{Ben{\'\i}tez-Llambay} \& {Masset}(2016)}]{Benitez2016}
{Ben{\'\i}tez-Llambay}, P. \& {Masset}, F.~S. 2016, \apjs, 223, 11

\bibitem[{{Birnstiel} {et~al.}(2010){Birnstiel}, {Dullemond}, \&
  {Brauer}}]{birnstiel2010}
{Birnstiel}, T., {Dullemond}, C.~P., \& {Brauer}, F. 2010, \aap, 513, A79

\bibitem[{{Birnstiel} {et~al.}(2018){Birnstiel}, {Dullemond}, {Zhu}, {Andrews},
  {Bai}, {Wilner}, {Carpenter}, {Huang}, {Isella}, {Benisty}, {P{\'e}rez}, \&
  {Zhang}}]{birnstiel2018}
{Birnstiel}, T., {Dullemond}, C.~P., {Zhu}, Z., {et~al.} 2018, \apjl, 869, L45

\bibitem[{{Blum} {et~al.}(2000){Blum}, {Wurm}, {Kempf}, {Poppe}, {Klahr},
  {Kozasa}, {Rott}, {Henning}, {Dorschner}, {Schr{\"a}pler}, {Keller},
  {Markiewicz}, {Mann}, {Gustafson}, {Giovane}, {Neuhaus}, {Fechtig},
  {Gr{\"u}n}, {Feuerbacher}, {Kochan}, {Ratke}, {El Goresy}, {Morfill},
  {Weidenschilling}, {Schwehm}, {Metzler}, \& {Ip}}]{blum2000}
{Blum}, J., {Wurm}, G., {Kempf}, S., {et~al.} 2000, \prl, 85, 2426

\bibitem[{{Campbell-White} {et~al.}(2023){Campbell-White}, {Manara}, {Benisty},
  {Natta}, {Claes}, {Frasca}, {Bae}, {Facchini}, {Isella}, {P{\'e}rez},
  {Pinilla}, {Sicilia-Aguilar}, \& {Teague}}]{Campbell2023}
{Campbell-White}, J., {Manara}, C.~F., {Benisty}, M., {et~al.} 2023, \apj, 956,
  25

\bibitem[{{Carrera} {et~al.}(2017){Carrera}, {Gorti}, {Johansen}, \&
  {Davies}}]{carrera2017}
{Carrera}, D., {Gorti}, U., {Johansen}, A., \& {Davies}, M.~B. 2017, \apj, 839,
  16

\bibitem[{{Christiaens} {et~al.}(2019){Christiaens}, {Casassus}, {Absil},
  {Cantalloube}, {Gomez Gonzalez}, {Girard}, {Ram{\'\i}rez}, {Pairet},
  {Salinas}, {Price}, {Pinte}, {Quanz}, {Jord{\'a}n}, {Mawet}, \&
  {Wahhaj}}]{Christiaens2019}
{Christiaens}, V., {Casassus}, S., {Absil}, O., {et~al.} 2019, \mnras, 486,
  5819

\bibitem[{{de Juan Ovelar} {et~al.}(2016){de Juan Ovelar}, {Pinilla}, {Min},
  {Dominik}, \& {Birnstiel}}]{ovelar2016}
{de Juan Ovelar}, M., {Pinilla}, P., {Min}, M., {Dominik}, C., \& {Birnstiel},
  T. 2016, \mnras, 459, L85

\bibitem[{{Delage} {et~al.}(2021){Delage}, {Okuzumi}, {Flock}, {Pinilla}, \&
  {Dzyurkevich}}]{delage2021}
{Delage}, T.~N., {Okuzumi}, S., {Flock}, M., {Pinilla}, P., \& {Dzyurkevich},
  N. 2021, arXiv e-prints, arXiv:2110.05639

\bibitem[{{Dominik} {et~al.}(2021){Dominik}, {Min}, \& {Tazaki}}]{dominik2021}
{Dominik}, C., {Min}, M., \& {Tazaki}, R. 2021, {OpTool: Command-line driven
  tool for creating complex dust opacities}

\bibitem[{{Dong} {et~al.}(2012){Dong}, {Hashimoto}, {Rafikov}, {Zhu},
  {Whitney}, {Kudo}, {Muto}, {Brandt}, {McClure}, {Wisniewski}, {Abe},
  {Brandner}, {Carson}, {Egner}, {Feldt}, {Goto}, {Grady}, {Guyon}, {Hayano},
  {Hayashi}, {Hayashi}, {Henning}, {Hodapp}, {Ishii}, {Iye}, {Janson},
  {Kandori}, {Knapp}, {Kusakabe}, {Kuzuhara}, {Kwon}, {Matsuo}, {McElwain},
  {Miyama}, {Morino}, {Moro-Martin}, {Nishimura}, {Pyo}, {Serabyn}, {Suto},
  {Suzuki}, {Takami}, {Takato}, {Terada}, {Thalmann}, {Tomono}, {Turner},
  {Watanabe}, {Yamada}, {Takami}, {Usuda}, \& {Tamura}}]{dong2012}
{Dong}, R., {Hashimoto}, J., {Rafikov}, R., {et~al.} 2012, \apj, 760, 111

\bibitem[{{Dong} {et~al.}(2017){Dong}, {Li}, {Chiang}, \& {Li}}]{dong2017b}
{Dong}, R., {Li}, S., {Chiang}, E., \& {Li}, H. 2017, \apj, 843, 127

\bibitem[{{Dr{\k{a}}{\.z}kowska} \& {Alibert}(2017)}]{joanna2017}
{Dr{\k{a}}{\.z}kowska}, J. \& {Alibert}, Y. 2017, \aap, 608, A92

\bibitem[{{Dr{\k{a}}{\.z}kowska} {et~al.}(2019){Dr{\k{a}}{\.z}kowska}, {Li},
  {Birnstiel}, {Stammler}, \& {Li}}]{joanna2019}
{Dr{\k{a}}{\.z}kowska}, J., {Li}, S., {Birnstiel}, T., {Stammler}, S.~M., \&
  {Li}, H. 2019, \apj, 885, 91

\bibitem[{{Dr{\k{a}}{\.z}kowska} \& {Szul{\'a}gyi}(2018)}]{joanna2018}
{Dr{\k{a}}{\.z}kowska}, J. \& {Szul{\'a}gyi}, J. 2018, \apj, 866, 142

\bibitem[{{Dullemond} {et~al.}(2012){Dullemond}, {Juhasz}, {Pohl}, {Sereshti},
  {Shetty}, {Peters}, {Commercon}, \& {Flock}}]{dullemond2012}
{Dullemond}, C.~P., {Juhasz}, A., {Pohl}, A., {et~al.} 2012, {RADMC-3D: A
  multi-purpose radiative transfer tool}

\bibitem[{{Facchini} {et~al.}(2021){Facchini}, {Teague}, {Bae}, {Benisty},
  {Keppler}, \& {Isella}}]{facchini2021}
{Facchini}, S., {Teague}, R., {Bae}, J., {et~al.} 2021, \aj, 162, 99

\bibitem[{{Gaia Collaboration} {et~al.}(2021){Gaia Collaboration}, {Brown},
  {Vallenari}, {Prusti}, {de Bruijne}, {Babusiaux}, {Biermann}, {Creevey},
  {Evans}, {Eyer}, {Hutton}, {Jansen}, {Jordi}, {Klioner}, {Lammers},
  {Lindegren}, {Luri}, {Mignard}, {Panem}, {Pourbaix}, {Randich}, {Sartoretti},
  {Soubiran}, {Walton}, {Arenou}, {Bailer-Jones}, {Bastian}, {Cropper},
  {Drimmel}, {Katz}, {Lattanzi}, {van Leeuwen}, {Bakker}, {Cacciari},
  {Casta{\~n}eda}, {De Angeli}, {Ducourant}, {Fabricius}, {Fouesneau},
  {Fr{\'e}mat}, {Guerra}, {Guerrier}, {Guiraud}, {Jean-Antoine Piccolo},
  {Masana}, {Messineo}, {Mowlavi}, {Nicolas}, {Nienartowicz}, {Pailler},
  {Panuzzo}, {Riclet}, {Roux}, {Seabroke}, {Sordo}, {Tanga}, {Th{\'e}venin},
  {Gracia-Abril}, {Portell}, {Teyssier}, {Altmann}, {Andrae}, {Bellas-Velidis},
  {Benson}, {Berthier}, {Blomme}, {Brugaletta}, {Burgess}, {Busso}, {Carry},
  {Cellino}, {Cheek}, {Clementini}, {Damerdji}, {Davidson}, {Delchambre},
  {Dell'Oro}, {Fern{\'a}ndez-Hern{\'a}ndez}, {Galluccio}, {Garc{\'\i}a-Lario},
  {Garcia-Reinaldos}, {Gonz{\'a}lez-N{\'u}{\~n}ez}, {Gosset}, {Haigron},
  {Halbwachs}, {Hambly}, {Harrison}, {Hatzidimitriou}, {Heiter},
  {Hern{\'a}ndez}, {Hestroffer}, {Hodgkin}, {Holl}, {Jan{\ss}en}, {Jevardat de
  Fombelle}, {Jordan}, {Krone-Martins}, {Lanzafame}, {L{\"o}ffler}, {Lorca},
  {Manteiga}, {Marchal}, {Marrese}, {Moitinho}, {Mora}, {Muinonen}, {Osborne},
  {Pancino}, {Pauwels}, {Petit}, {Recio-Blanco}, {Richards}, {Riello},
  {Rimoldini}, {Robin}, {Roegiers}, {Rybizki}, {Sarro}, {Siopis}, {Smith},
  {Sozzetti}, {Ulla}, {Utrilla}, {van Leeuwen}, {van Reeven}, {Abbas}, {Abreu
  Aramburu}, {Accart}, {Aerts}, {Aguado}, {Ajaj}, {Altavilla}, {{\'A}lvarez},
  {{\'A}lvarez Cid-Fuentes}, {Alves}, {Anderson}, {Anglada Varela}, {Antoja},
  {Audard}, {Baines}, {Baker}, {Balaguer-N{\'u}{\~n}ez}, {Balbinot}, {Balog},
  {Barache}, {Barbato}, {Barros}, {Barstow}, {Bartolom{\'e}}, {Bassilana},
  {Bauchet}, {Baudesson-Stella}, {Becciani}, {Bellazzini}, {Bernet}, {Bertone},
  {Bianchi}, {Blanco-Cuaresma}, {Boch}, {Bombrun}, {Bossini}, {Bouquillon},
  {Bragaglia}, {Bramante}, {Breedt}, {Bressan}, {Brouillet}, {Bucciarelli},
  {Burlacu}, {Busonero}, {Butkevich}, {Buzzi}, {Caffau}, {Cancelliere},
  {C{\'a}novas}, {Cantat-Gaudin}, {Carballo}, {Carlucci}, {Carnerero},
  {Carrasco}, {Casamiquela}, {Castellani}, {Castro-Ginard}, {Castro Sampol},
  {Chaoul}, {Charlot}, {Chemin}, {Chiavassa}, {Cioni}, {Comoretto}, {Cooper},
  {Cornez}, {Cowell}, {Crifo}, {Crosta}, {Crowley}, {Dafonte}, {Dapergolas},
  {David}, {David}, {de Laverny}, {De Luise}, {De March}, {De Ridder}, {de
  Souza}, {de Teodoro}, {de Torres}, {del Peloso}, {del Pozo}, {Delbo},
  {Delgado}, {Delgado}, {Delisle}, {Di Matteo}, {Diakite}, {Diener},
  {Distefano}, {Dolding}, {Eappachen}, {Edvardsson}, {Enke}, {Esquej}, {Fabre},
  {Fabrizio}, {Faigler}, {Fedorets}, {Fernique}, {Fienga}, {Figueras},
  {Fouron}, {Fragkoudi}, {Fraile}, {Franke}, {Gai}, {Garabato},
  {Garcia-Gutierrez}, {Garc{\'\i}a-Torres}, {Garofalo}, {Gavras}, {Gerlach},
  {Geyer}, {Giacobbe}, {Gilmore}, {Girona}, {Giuffrida}, {Gomel}, {Gomez},
  {Gonzalez-Santamaria}, {Gonz{\'a}lez-Vidal}, {Granvik},
  {Guti{\'e}rrez-S{\'a}nchez}, {Guy}, {Hauser}, {Haywood}, {Helmi}, {Hidalgo},
  {Hilger}, {H{\l}adczuk}, {Hobbs}, {Holland}, {Huckle}, {Jasniewicz},
  {Jonker}, {Juaristi Campillo}, {Julbe}, {Karbevska}, {Kervella}, {Khanna},
  {Kochoska}, {Kontizas}, {Kordopatis}, {Korn}, {Kostrzewa-Rutkowska},
  {Kruszy{\'n}ska}, {Lambert}, {Lanza}, {Lasne}, {Le Campion}, {Le Fustec},
  {Lebreton}, {Lebzelter}, {Leccia}, {Leclerc}, {Lecoeur-Taibi}, {Liao},
  {Licata}, {Lindstr{\o}m}, {Lister}, {Livanou}, {Lobel}, {Madrero Pardo},
  {Managau}, {Mann}, {Marchant}, {Marconi}, {Marcos Santos}, {Marinoni},
  {Marocco}, {Marshall}, {Martin Polo}, {Mart{\'\i}n-Fleitas}, {Masip},
  {Massari}, {Mastrobuono-Battisti}, {Mazeh}, {McMillan}, {Messina},
  {Michalik}, {Millar}, {Mints}, {Molina}, {Molinaro}, {Moln{\'a}r},
  {Montegriffo}, {Mor}, {Morbidelli}, {Morel}, {Morris}, {Mulone}, {Munoz},
  {Muraveva}, {Murphy}, {Musella}, {Noval}, {Ord{\'e}novic}, {Orr{\`u}},
  {Osinde}, {Pagani}, {Pagano}, {Palaversa}, {Palicio}, {Panahi}, {Pawlak},
  {Pe{\~n}alosa Esteller}, {Penttil{\"a}}, {Piersimoni}, {Pineau}, {Plachy},
  {Plum}, {Poggio}, {Poretti}, {Poujoulet}, {Pr{\v{s}}a}, {Pulone}, {Racero},
  {Ragaini}, {Rainer}, {Raiteri}, {Rambaux}, {Ramos}, {Ramos-Lerate}, {Re
  Fiorentin}, {Regibo}, {Reyl{\'e}}, {Ripepi}, {Riva}, {Rixon}, {Robichon},
  {Robin}, {Roelens}, {Rohrbasser}, {Romero-G{\'o}mez}, {Rowell}, {Royer},
  {Rybicki}, {Sadowski}, {Sagrist{\`a} Sell{\'e}s}, {Sahlmann}, {Salgado},
  {Salguero}, {Samaras}, {Sanchez Gimenez}, {Sanna}, {Santove{\~n}a},
  {Sarasso}, {Schultheis}, {Sciacca}, {Segol}, {Segovia}, {S{\'e}gransan},
  {Semeux}, {Shahaf}, {Siddiqui}, {Siebert}, {Siltala}, {Slezak}, {Smart},
  {Solano}, {Solitro}, {Souami}, {Souchay}, {Spagna}, {Spoto}, {Steele},
  {Steidelm{\"u}ller}, {Stephenson}, {S{\"u}veges}, {Szabados}, {Szegedi-Elek},
  {Taris}, {Tauran}, {Taylor}, {Teixeira}, {Thuillot}, {Tonello}, {Torra},
  {Torra}, {Turon}, {Unger}, {Vaillant}, {van Dillen}, {Vanel}, {Vecchiato},
  {Viala}, {Vicente}, {Voutsinas}, {Weiler}, {Wevers}, {Wyrzykowski}, {Yoldas},
  {Yvard}, {Zhao}, {Zorec}, {Zucker}, {Zurbach}, \& {Zwitter}}]{gaia2021edr3}
{Gaia Collaboration}, {Brown}, A.~G.~A., {Vallenari}, A., {et~al.} 2021, \aap,
  649, A1

\bibitem[{{Gundlach} {et~al.}(2018){Gundlach}, {Schmidt}, {Kreuzig},
  {Bischoff}, {Rezaei}, {Kothe}, {Blum}, {Grzesik}, \& {Stoll}}]{gundlach2018}
{Gundlach}, B., {Schmidt}, K.~P., {Kreuzig}, C., {et~al.} 2018, \mnras, 479,
  1273

\bibitem[{{Habart} {et~al.}(2021){Habart}, {Bout{\'e}raon}, {Brauer}, {Ysard},
  {Pantin}, {Marchal}, \& {Jones}}]{habart2021}
{Habart}, E., {Bout{\'e}raon}, T., {Brauer}, R., {et~al.} 2021, \aap, 649, A84

\bibitem[{{Haffert} {et~al.}(2019){Haffert}, {Bohn}, {de Boer}, {Snellen},
  {Brinchmann}, {Girard}, {Keller}, \& {Bacon}}]{haffert2019}
{Haffert}, S.~Y., {Bohn}, A.~J., {de Boer}, J., {et~al.} 2019, Nature
  Astronomy, 3, 749

\bibitem[{{Hashimoto} {et~al.}(2020){Hashimoto}, {Aoyama}, {Konishi}, {Uyama},
  {Takasao}, {Ikoma}, \& {Tanigawa}}]{hashimoto2020}
{Hashimoto}, J., {Aoyama}, Y., {Konishi}, M., {et~al.} 2020, \aj, 159, 222

\bibitem[{{Hashimoto} {et~al.}(2015){Hashimoto}, {Tsukagoshi}, {Brown}, {Dong},
  {Muto}, {Zhu}, {Wisniewski}, {Ohashi}, {kudo}, {Kusakabe}, {Abe}, {Akiyama},
  {Brandner}, {Brandt}, {Carson}, {Currie}, {Egner}, {Feldt}, {Grady}, {Guyon},
  {Hayano}, {Hayashi}, {Hayashi}, {Henning}, {Hodapp}, {Ishii}, {Iye},
  {Janson}, {Kandori}, {Knapp}, {Kuzuhara}, {Kwon}, {Matsuo}, {McElwain},
  {Mayama}, {Mede}, {Miyama}, {Morino}, {Moro-Martin}, {Nishimura}, {Pyo},
  {Serabyn}, {Suenaga}, {Suto}, {Suzuki}, {Takahashi}, {Takami}, {Takato},
  {Terada}, {Thalmann}, {Tomono}, {Turner}, {Watanabe}, {Yamada}, {Takami},
  {Usuda}, \& {Tamura}}]{hashimoto2015}
{Hashimoto}, J., {Tsukagoshi}, T., {Brown}, J.~M., {et~al.} 2015, \apj, 799, 43

\bibitem[{{Heays} {et~al.}(2017){Heays}, {Bosman}, \& {van
  Dishoeck}}]{heays2017}
{Heays}, A.~N., {Bosman}, A.~D., \& {van Dishoeck}, E.~F. 2017, \aap, 602, A105

\bibitem[{{Hensley} \& {Draine}(2023)}]{hensley2023}
{Hensley}, B.~S. \& {Draine}, B.~T. 2023, \apj, 948, 55

\bibitem[{{Isella} {et~al.}(2019){Isella}, {Benisty}, {Teague}, {Bae},
  {Keppler}, {Facchini}, \& {P{\'e}rez}}]{isella2019}
{Isella}, A., {Benisty}, M., {Teague}, R., {et~al.} 2019, \apjl, 879, L25

\bibitem[{{Jones} {et~al.}(2017){Jones}, {K{\"o}hler}, {Ysard}, {Bocchio}, \&
  {Verstraete}}]{jones2017}
{Jones}, A.~P., {K{\"o}hler}, M., {Ysard}, N., {Bocchio}, M., \& {Verstraete},
  L. 2017, \aap, 602, A46

\bibitem[{{Kalyaan} \& {Desch}(2019)}]{kalyaan2019}
{Kalyaan}, A. \& {Desch}, S.~J. 2019, \apj, 875, 43

\bibitem[{{Kalyaan} {et~al.}(2023){Kalyaan}, {Pinilla}, {Krijt}, {Banzatti},
  {Rosotti}, {Mulders}, {Lambrechts}, {Long}, \& {Herczeg}}]{kalyaan2023}
{Kalyaan}, A., {Pinilla}, P., {Krijt}, S., {et~al.} 2023, \apj, 954, 66

\bibitem[{{Kalyaan} {et~al.}(2021){Kalyaan}, {Pinilla}, {Krijt}, {Mulders}, \&
  {Banzatti}}]{kalyaan2021}
{Kalyaan}, A., {Pinilla}, P., {Krijt}, S., {Mulders}, G.~D., \& {Banzatti}, A.
  2021, \apj, 921, 84

\bibitem[{{Keppler} {et~al.}(2018){Keppler}, {Benisty}, {M{\"u}ller},
  {Henning}, {van Boekel}, {Cantalloube}, {Ginski}, {van Holstein}, {Maire},
  {Pohl}, {Samland}, {Avenhaus}, {Baudino}, {Boccaletti}, {de Boer},
  {Bonnefoy}, {Chauvin}, {Desidera}, {Langlois}, {Lazzoni}, {Marleau},
  {Mordasini}, {Pawellek}, {Stolker}, {Vigan}, {Zurlo}, {Birnstiel},
  {Brandner}, {Feldt}, {Flock}, {Girard}, {Gratton}, {Hagelberg}, {Isella},
  {Janson}, {Juhasz}, {Kemmer}, {Kral}, {Lagrange}, {Launhardt}, {Matter},
  {M{\'e}nard}, {Milli}, {Molli{\`e}re}, {Olofsson}, {P{\'e}rez}, {Pinilla},
  {Pinte}, {Quanz}, {Schmidt}, {Udry}, {Wahhaj}, {Williams}, {Buenzli},
  {Cudel}, {Dominik}, {Galicher}, {Kasper}, {Lannier}, {Mesa}, {Mouillet},
  {Peretti}, {Perrot}, {Salter}, {Sissa}, {Wildi}, {Abe}, {Antichi},
  {Augereau}, {Baruffolo}, {Baudoz}, {Bazzon}, {Beuzit}, {Blanchard}, {Brems},
  {Buey}, {De Caprio}, {Carbillet}, {Carle}, {Cascone}, {Cheetham}, {Claudi},
  {Costille}, {Delboulb{\'e}}, {Dohlen}, {Fantinel}, {Feautrier}, {Fusco},
  {Giro}, {Gluck}, {Gry}, {Hubin}, {Hugot}, {Jaquet}, {Le Mignant}, {Llored},
  {Madec}, {Magnard}, {Martinez}, {Maurel}, {Meyer}, {M{\"o}ller-Nilsson},
  {Moulin}, {Mugnier}, {Orign{\'e}}, {Pavlov}, {Perret}, {Petit}, {Pragt},
  {Puget}, {Rabou}, {Ramos}, {Rigal}, {Rochat}, {Roelfsema}, {Rousset}, {Roux},
  {Salasnich}, {Sauvage}, {Sevin}, {Soenke}, {Stadler}, {Suarez}, {Turatto}, \&
  {Weber}}]{keppler2018}
{Keppler}, M., {Benisty}, M., {M{\"u}ller}, A., {et~al.} 2018, \aap, 617, A44

\bibitem[{{Keppler} {et~al.}(2019){Keppler}, {Teague}, {Bae}, {Benisty},
  {Henning}, {van Boekel}, {Chapillon}, {Pinilla}, {Williams}, {Bertrang},
  {Facchini}, {Flock}, {Ginski}, {Juhasz}, {Klahr}, {Liu}, {M{\"u}ller},
  {P{\'e}rez}, {Pohl}, {Rosotti}, {Samland}, \& {Semenov}}]{keppler2019}
{Keppler}, M., {Teague}, R., {Bae}, J., {et~al.} 2019, \aap, 625, A118

\bibitem[{{K{\"o}hler} {et~al.}(2015){K{\"o}hler}, {Ysard}, \&
  {Jones}}]{koehler2015}
{K{\"o}hler}, M., {Ysard}, N., \& {Jones}, A.~P. 2015, \aap, 579, A15

\bibitem[{{Kokoulina} {et~al.}(2021){Kokoulina}, {Matter}, {Lopez}, {Pantin},
  {Ysard}, {Weigelt}, {Habart}, {Varga}, {Jones}, {Meilland}, {Dartois},
  {Klarmann}, {Augereau}, {van Boekel}, {Hogerheijde}, {Yoffe}, {Waters},
  {Dominik}, {Jaffe}, {Millour}, {Henning}, {Hofmann}, {Schertl}, {Lagarde},
  {Petrov}, {Antonelli}, {Allouche}, {Berio}, {Robbe-Dubois}, {{\'A}braham},
  {Beckmann}, {Bensberg}, {Bettonvil}, {Bristow}, {Cruzal{\`e}bes}, {Danchi},
  {Dannhoff}, {Graser}, {Heininger}, {Labadie}, {Lehmitz}, {Leinert},
  {Meisenheimer}, {Paladini}, {Percheron}, {Stee}, {Woillez}, {Wolf}, {Zins},
  {Delbo}, {Drevon}, {Duprat}, {G{\'a}mez Rosas}, {Hocd{\'e}}, {Hron},
  {Hummel}, {Isbell}, {Leftley}, {Soulain}, {Vakili}, \&
  {Wittkowski}}]{Kokoulina2021}
{Kokoulina}, E., {Matter}, A., {Lopez}, B., {et~al.} 2021, \aap, 652, A61

\bibitem[{{Krijt} {et~al.}(2020){Krijt}, {Bosman}, {Zhang}, {Schwarz},
  {Ciesla}, \& {Bergin}}]{krijt2020}
{Krijt}, S., {Bosman}, A.~D., {Zhang}, K., {et~al.} 2020, \apj, 899, 134

\bibitem[{{Lesur} {et~al.}(2023){Lesur}, {Flock}, {Ercolano}, {Lin}, {Yang},
  {Barranco}, {Benitez-Llambay}, {Goodman}, {Johansen}, {Klahr}, {Laibe},
  {Lyra}, {Marcus}, {Nelson}, {Squire}, {Simon}, {Turner}, {Umurhan}, \&
  {Youdin}}]{lesur2023}
{Lesur}, G., {Flock}, M., {Ercolano}, B., {et~al.} 2023, in Astronomical
  Society of the Pacific Conference Series, Vol. 534, Astronomical Society of
  the Pacific Conference Series, ed. S.~{Inutsuka}, Y.~{Aikawa}, T.~{Muto},
  K.~{Tomida}, \& M.~{Tamura}, 465

\bibitem[{{Long} {et~al.}(2018){Long}, {Akiyama}, {Sitko}, {Fernandes},
  {Assani}, {Grady}, {Cure}, {Danchi}, {Dong}, {Fukagawa}, {Hasegawa},
  {Hashimoto}, {Henning}, {Inutsuka}, {Kraus}, {Kwon}, {Lisse}, {Liu},
  {Mayama}, {Muto}, {Nakagawa}, {Takami}, {Tamura}, {Currie}, {Wisniewski}, \&
  {Yang}}]{long2018_pds}
{Long}, Z.~C., {Akiyama}, E., {Sitko}, M., {et~al.} 2018, \apj, 858, 112

\bibitem[{{Mathis} {et~al.}(1977){Mathis}, {Rumpl}, \&
  {Nordsieck}}]{mathis1977}
{Mathis}, J.~S., {Rumpl}, W., \& {Nordsieck}, K.~H. 1977, \apj, 217, 425

\bibitem[{{M{\"u}ller} {et~al.}(2018){M{\"u}ller}, {Keppler}, {Henning},
  {Samland}, {Chauvin}, {Beust}, {Maire}, {Molaverdikhani}, {van Boekel},
  {Benisty}, {Boccaletti}, {Bonnefoy}, {Cantalloube}, {Charnay}, {Baudino},
  {Gennaro}, {Long}, {Cheetham}, {Desidera}, {Feldt}, {Fusco}, {Girard},
  {Gratton}, {Hagelberg}, {Janson}, {Lagrange}, {Langlois}, {Lazzoni}, {Ligi},
  {M{\'e}nard}, {Mesa}, {Meyer}, {Molli{\`e}re}, {Mordasini}, {Moulin},
  {Pavlov}, {Pawellek}, {Quanz}, {Ramos}, {Rouan}, {Sissa}, {Stadler}, {Vigan},
  {Wahhaj}, {Weber}, \& {Zurlo}}]{muller2018}
{M{\"u}ller}, A., {Keppler}, M., {Henning}, T., {et~al.} 2018, \aap, 617, L2

\bibitem[{{Musiolik} \& {Wurm}(2019)}]{musiolik2019}
{Musiolik}, G. \& {Wurm}, G. 2019, \apj, 873, 58

\bibitem[{{Perotti} {et~al.}(2023){Perotti}, {Christiaens}, {Henning},
  {Tabone}, {Waters}, {Kamp}, {Olofsson}, {Grant}, {Gasman}, {Bouwman},
  {Samland}, {Franceschi}, {van Dishoeck}, {Schwarz}, {G{\"u}del}, {Lagage},
  {Ray}, {Vandenbussche}, {Abergel}, {Absil}, {Arabhavi}, {Argyriou},
  {Barrado}, {Boccaletti}, {Caratti o Garatti}, {Geers}, {Glauser},
  {Justannont}, {Lahuis}, {Mueller}, {Nehm{\'e}}, {Pantin}, {Scheithauer},
  {Waelkens}, {Guadarrama}, {Jang}, {Kanwar}, {Morales-Calder{\'o}n},
  {Pawellek}, {Rodgers-Lee}, {Schreiber}, {Colina}, {Greve}, {{\"O}stlin}, \&
  {Wright}}]{perotti2023}
{Perotti}, G., {Christiaens}, V., {Henning}, T., {et~al.} 2023, \nat, 620, 516

\bibitem[{{Pinilla} {et~al.}(2012){Pinilla}, {Benisty}, \&
  {Birnstiel}}]{pinilla2012}
{Pinilla}, P., {Benisty}, M., \& {Birnstiel}, T. 2012, \aap, 545, A81

\bibitem[{{Pinilla} {et~al.}(2014){Pinilla}, {Benisty}, {Birnstiel}, {Ricci},
  {Isella}, {Natta}, {Dullemond}, {Quiroga-Nu{\~n}ez}, {Henning}, \&
  {Testi}}]{pinilla2014}
{Pinilla}, P., {Benisty}, M., {Birnstiel}, T., {et~al.} 2014, \aap, 564, A51

\bibitem[{{Pinilla} {et~al.}(2016){Pinilla}, {Klarmann}, {Birnstiel},
  {Benisty}, {Dominik}, \& {Dullemond}}]{pinilla2016}
{Pinilla}, P., {Klarmann}, L., {Birnstiel}, T., {et~al.} 2016, \aap, 585, A35

\bibitem[{{Pinilla} {et~al.}(2021){Pinilla}, {Lenz}, \&
  {Stammler}}]{pinilla2021}
{Pinilla}, P., {Lenz}, C.~T., \& {Stammler}, S.~M. 2021, \aap, 645, A70

\bibitem[{{Pohl} {et~al.}(2017){Pohl}, {Benisty}, {Pinilla}, {Ginski}, {de
  Boer}, {Avenhaus}, {Henning}, {Zurlo}, {Boccaletti}, {Augereau}, {Birnstiel},
  {Dominik}, {Facchini}, {Fedele}, {Janson}, {Keppler}, {Kral}, {Langlois},
  {Ligi}, {Maire}, {M{\'e}nard}, {Meyer}, {Pinte}, {Quanz}, {Sauvage},
  {Sezestre}, {Stolker}, {Szul{\'a}gyi}, {van Boekel}, {van der Plas},
  {Villenave}, {Baruffolo}, {Baudoz}, {Le Mignant}, {Maurel}, {Ramos}, \&
  {Weber}}]{pohl2017}
{Pohl}, A., {Benisty}, M., {Pinilla}, P., {et~al.} 2017, \apj, 850, 52

\bibitem[{{Portilla-Revelo} {et~al.}(2023){Portilla-Revelo}, {Kamp},
  {Facchini}, {van Dishoeck}, {Law}, {Rab}, {Bae}, {Benisty}, {{\"O}berg}, \&
  {Teague}}]{portilla2023}
{Portilla-Revelo}, B., {Kamp}, I., {Facchini}, S., {et~al.} 2023, \aap, 677,
  A76

\bibitem[{{Ren} {et~al.}(2023){Ren}, {Benisty}, {Ginski}, {Tazaki}, {Wallack},
  {Milli}, {Garufi}, {Bae}, {Facchini}, {M{\'e}nard}, {Pinilla}, {Swastik},
  {Teague}, \& {Wahhaj}}]{Ren2023}
{Ren}, B.~B., {Benisty}, M., {Ginski}, C., {et~al.} 2023, arXiv e-prints,
  arXiv:2310.08589

\bibitem[{{Ricci} {et~al.}(2010){Ricci}, {Testi}, {Natta}, {Neri}, {Cabrit}, \&
  {Herczeg}}]{ricci2010}
{Ricci}, L., {Testi}, L., {Natta}, A., {et~al.} 2010, \aap, 512, A15

\bibitem[{{Rosotti}(2023)}]{rosotti2023}
{Rosotti}, G.~P. 2023, \nar, 96, 101674

\bibitem[{{Salyk} {et~al.}(2019){Salyk}, {Lacy}, {Richter}, {Zhang},
  {Pontoppidan}, {Carr}, {Najita}, \& {Blake}}]{salyk2019}
{Salyk}, C., {Lacy}, J., {Richter}, M., {et~al.} 2019, \apj, 874, 24

\bibitem[{{Salyk} {et~al.}(2011){Salyk}, {Pontoppidan}, {Blake}, {Najita}, \&
  {Carr}}]{salyk2011}
{Salyk}, C., {Pontoppidan}, K.~M., {Blake}, G.~A., {Najita}, J.~R., \& {Carr},
  J.~S. 2011, \apj, 731, 130

\bibitem[{{Schwarz} {et~al.}(2018){Schwarz}, {Bergin}, {Cleeves}, {Zhang},
  {{\"O}berg}, {Blake}, \& {Anderson}}]{schwarz2018}
{Schwarz}, K.~R., {Bergin}, E.~A., {Cleeves}, L.~I., {et~al.} 2018, \apj, 856,
  85

\bibitem[{{Stadler} {et~al.}(2022){Stadler}, {G{\'a}rate}, {Pinilla}, {Lenz},
  {Dullemond}, {Birnstiel}, \& {Stammler}}]{stadler2022}
{Stadler}, J., {G{\'a}rate}, M., {Pinilla}, P., {et~al.} 2022, \aap, 668, A104

\bibitem[{{Stammler} \& {Birnstiel}(2022)}]{stammler2022}
{Stammler}, S.~M. \& {Birnstiel}, T. 2022, \apj, 935, 35

\bibitem[{{Steinpilz} {et~al.}(2019){Steinpilz}, {Teiser}, \&
  {Wurm}}]{steinpilz2019}
{Steinpilz}, T., {Teiser}, J., \& {Wurm}, G. 2019, \apj, 874, 60

\bibitem[{{Tazzari} {et~al.}(2021){Tazzari}, {Testi}, {Natta}, {Williams},
  {Ansdell}, {Carpenter}, {Facchini}, {Guidi}, {Hogherheijde}, {Manara},
  {Miotello}, \& {van der Marel}}]{tazzari2021}
{Tazzari}, M., {Testi}, L., {Natta}, A., {et~al.} 2021, \mnras, 506, 5117

\bibitem[{{Thanathibodee} {et~al.}(2019){Thanathibodee}, {Calvet}, {Bae},
  {Muzerolle}, \& {Hern{\'a}ndez}}]{Thanathibodee2020}
{Thanathibodee}, T., {Calvet}, N., {Bae}, J., {Muzerolle}, J., \&
  {Hern{\'a}ndez}, R.~F. 2019, \apj, 885, 94

\bibitem[{{Toci} {et~al.}(2020){Toci}, {Lodato}, {Christiaens}, {Fedele},
  {Pinte}, {Price}, \& {Testi}}]{toci2020}
{Toci}, C., {Lodato}, G., {Christiaens}, V., {et~al.} 2020, \mnras, 499, 2015

\bibitem[{{Trapman} {et~al.}(2022){Trapman}, {Zhang}, {van't Hoff},
  {Hogerheijde}, \& {Bergin}}]{trapman2022}
{Trapman}, L., {Zhang}, K., {van't Hoff}, M. L.~R., {Hogerheijde}, M.~R., \&
  {Bergin}, E.~A. 2022, \apjl, 926, L2

\bibitem[{{Virtanen} {et~al.}(2020){Virtanen}, {Gommers}, {Oliphant},
  {Haberland}, {Reddy}, {Cournapeau}, {Burovski}, {Peterson}, {Weckesser},
  {Bright}, {van der Walt}, {Brett}, {Wilson}, {Millman}, {Mayorov}, {Nelson},
  {Jones}, {Kern}, {Larson}, {Carey}, {Polat}, {Feng}, {Moore}, {VanderPlas},
  {Laxalde}, {Perktold}, {Cimrman}, {Henriksen}, {Quintero}, {Harris},
  {Archibald}, {Ribeiro}, {Pedregosa}, {van Mulbregt}, \& {SciPy 1. 0
  Contributors}}]{virtanen2020}
{Virtanen}, P., {Gommers}, R., {Oliphant}, T.~E., {et~al.} 2020, Nature
  Methods, 17, 261

\bibitem[{{Wada} {et~al.}(2009){Wada}, {Tanaka}, {Suyama}, {Kimura}, \&
  {Yamamoto}}]{wada2009}
{Wada}, K., {Tanaka}, H., {Suyama}, T., {Kimura}, H., \& {Yamamoto}, T. 2009,
  \apj, 702, 1490

\bibitem[{{Wang} {et~al.}(2021){Wang}, {Vigan}, {Lacour}, {Nowak}, {Stolker},
  {De Rosa}, {Ginzburg}, {Gao}, {Abuter}, {Amorim}, {Asensio-Torres},
  {Baub{\"o}ck}, {Benisty}, {Berger}, {Beust}, {Beuzit}, {Blunt}, {Boccaletti},
  {Bohn}, {Bonnefoy}, {Bonnet}, {Brandner}, {Cantalloube}, {Caselli},
  {Charnay}, {Chauvin}, {Choquet}, {Christiaens}, {Cl{\'e}net}, {Coud{\'e} Du
  Foresto}, {Cridland}, {de Zeeuw}, {Dembet}, {Dexter}, {Drescher}, {Duvert},
  {Eckart}, {Eisenhauer}, {Facchini}, {Gao}, {Garcia}, {Garcia Lopez},
  {Gardner}, {Gendron}, {Genzel}, {Gillessen}, {Girard}, {Haubois},
  {Hei{\ss}el}, {Henning}, {Hinkley}, {Hippler}, {Horrobin}, {Houll{\'e}},
  {Hubert}, {Jim{\'e}nez-Rosales}, {Jocou}, {Kammerer}, {Keppler}, {Kervella},
  {Meyer}, {Kreidberg}, {Lagrange}, {Lapeyr{\`e}re}, {Le Bouquin}, {L{\'e}na},
  {Lutz}, {Maire}, {M{\'e}nard}, {M{\'e}rand}, {Molli{\`e}re}, {Monnier},
  {Mouillet}, {M{\"u}ller}, {Nasedkin}, {Ott}, {Otten}, {Paladini}, {Paumard},
  {Perraut}, {Perrin}, {Pfuhl}, {Pueyo}, {Rameau}, {Rodet},
  {Rodr{\'\i}guez-Coira}, {Rousset}, {Scheithauer}, {Shangguan}, {Shimizu},
  {Stadler}, {Straub}, {Straubmeier}, {Sturm}, {Tacconi}, {van Dishoeck},
  {Vincent}, {von Fellenberg}, {Ward-Duong}, {Widmann}, {Wieprecht},
  {Wiezorrek}, {Woillez}, \& {Gravity Collaboration}}]{wang2021}
{Wang}, J.~J., {Vigan}, A., {Lacour}, S., {et~al.} 2021, \aj, 161, 148

\bibitem[{{Youdin} \& {Lithwick}(2007)}]{youdin2007}
{Youdin}, A.~N. \& {Lithwick}, Y. 2007, \icarus, 192, 588

\bibitem[{{Zhou} {et~al.}(2021){Zhou}, {Bowler}, {Wagner}, {Schneider}, {Apai},
  {Kraus}, {Close}, {Herczeg}, \& {Fang}}]{zhou2021}
{Zhou}, Y., {Bowler}, B.~P., {Wagner}, K.~R., {et~al.} 2021, \aj, 161, 244

\bibitem[{{Zormpas} {et~al.}(2022){Zormpas}, {Birnstiel}, {Rosotti}, \&
  {Andrews}}]{zormpas2022}
{Zormpas}, A., {Birnstiel}, T., {Rosotti}, G.~P., \& {Andrews}, S.~M. 2022,
  \aap, 661, A66

\end{thebibliography}


\begin{appendix}

\section{Radial resolution test} \label{appendixA}

\begin{figure*}[h!]
\centering
    \tabcolsep=0.05cm 
    \begin{tabular}{cc}  
     \includegraphics[width=9cm]{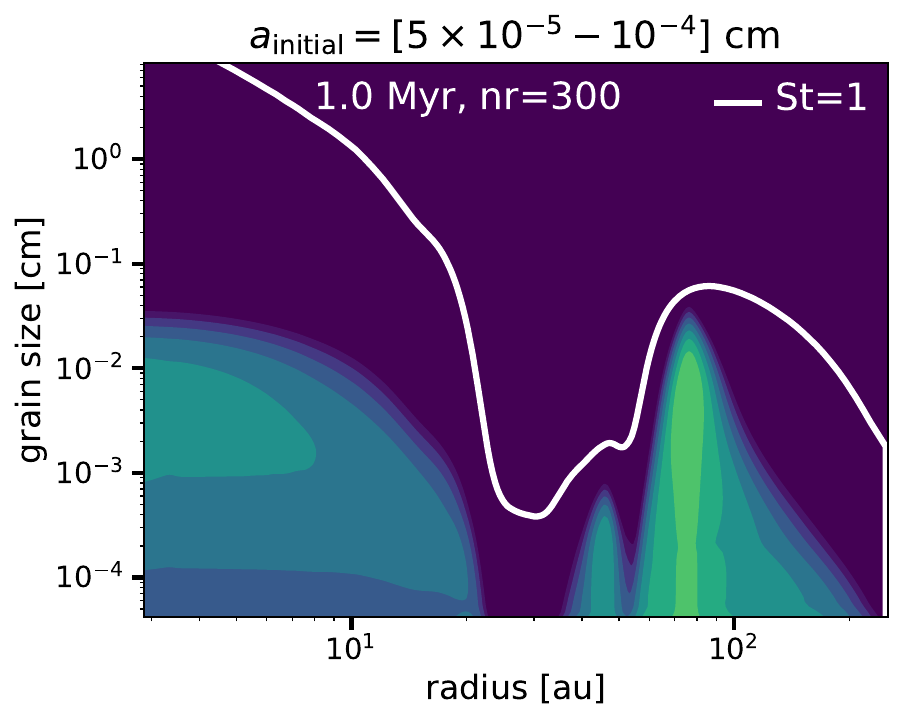}&
     \includegraphics[width=9cm]{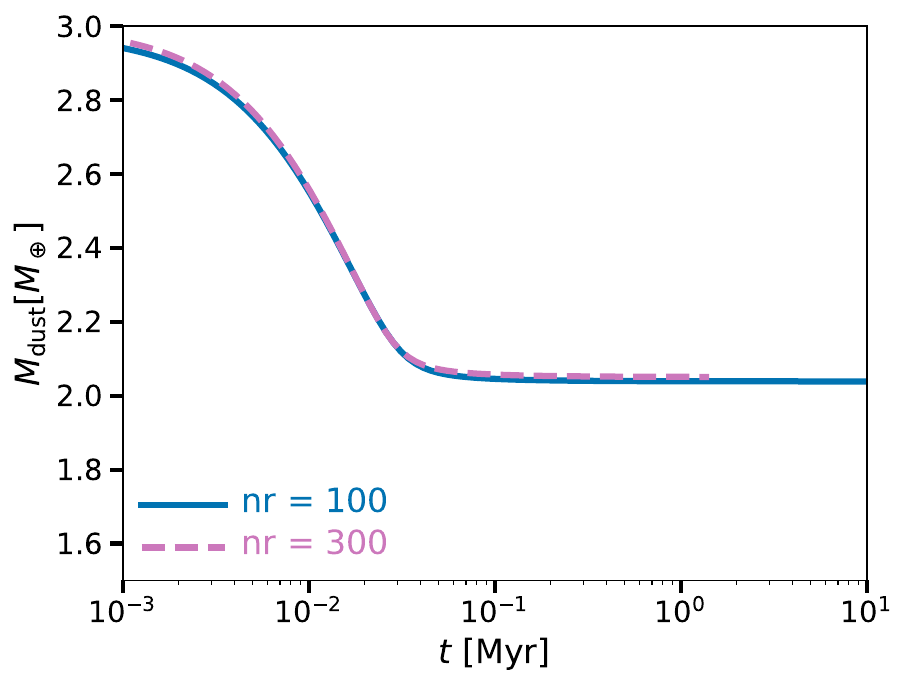}
    \end{tabular}
\caption{{Resolution test. Left panel: Dust density distribution for the \PDS~disk after 1\,Myr of evolution as in Fig.~\ref{fig:dust_evolution_models} when $a_{\rm{initial}}=[5\times10^{-5}-10^{-4}]$\,cm, and with a radial resolution of $n_r =300$. Right panel: Total dust disk mass as a function of time for the models considering $a_{\rm{initial}}=[5\times10^{-5}-10^{-4}]$\,cm and a radial resolution of the simulation of $n_r =300$ and $n_r =100$. }}
\label{fig:resolution_Test}
\end{figure*}

{We performed a dust evolution model to test the radial resolution of our simulations. The left panel of Fig.~\ref{fig:resolution_Test} shows the dust density distribution at 1\,Myr of evolution for the case of  $a_{\rm{initial}}=[5\times10^{-5}-10^{-4}]$\,cm as in the top left panel of Fig.~\ref{fig:dust_evolution_models}, but with a higher number of radial cells in the simulations ($n_r=300$). The right panel of Fig.~\ref{fig:resolution_Test} shows the total dust disk mass as a function of time for both values of the radial resolution ($n_r =300$ and $n_r =100$; due to the computational cost of the high-resolution simulations, we only show results up to 1.5\,Myr after the dust disk mass no longer changes over time). Figure~\ref{fig:resolution_Test} shows no differences in the results when assuming $n_r =300$ versus $n_r =100$ }

\end{appendix}

\end{document}